\pdfoutput=1
\documentclass{article}
\usepackage{arxiv}

\usepackage[utf8]{inputenc}
\usepackage[T1]{fontenc}

\usepackage{amsmath,amssymb,amsfonts}
\usepackage{mathtools}

\usepackage{graphicx}
\graphicspath{{./images/}}

\usepackage{booktabs}
\usepackage{array}
\usepackage{makecell}
\usepackage{threeparttable}
\usepackage{dcolumn}

\usepackage{textcomp}
\usepackage{nicefrac}
\usepackage{microtype}
\usepackage{times}

\usepackage[table]{xcolor}
\usepackage{soul}
\colorlet{blue}{blue}

\usepackage[numbers,sort&compress]{natbib}

\usepackage{hyperref}
\usepackage{url}
\hypersetup{
  colorlinks=true,
  linkcolor=blue,
  filecolor=magenta,
  urlcolor=cyan,
}

\usepackage{caption}
\usepackage{subcaption}

\usepackage{multicol}
\usepackage{balance}
\usepackage{flushend}
\usepackage{afterpage}
\usepackage{float}

\usepackage[ruled,vlined,linesnumbered]{algorithm2e}
\usepackage{enumitem}

\usepackage{listings}
\usepackage{courier}

\usepackage{comment}

\title{Stringology-Based Cryptanalysis for EChaCha20 Stream Cipher}

\author{
 Victor Kebande \\
  Department of Coumputing and Engineering\\
  University of Colorado Denver\\
  380, Lawrence St, Denver, 802024, USA \\
  \texttt{victor.kebande@ucdenver.edu} \\
}

\begin{document}
\maketitle
\begin{abstract}
Stringology-Based Cryptanalysis (SBC) offers a suitable and a structurally aligned approach for uncovering
structural patterns in stream ciphers that traditional statistical tests may often fail to detect. 
Despite \texttt{EChaCha20}'s design enhancements, no systematic investigation has been performed to determine whether its expanded 6$\times$6 state matrix and modified Quarter-Round Function (\texttt{QR-F}) introduce subtle keystream patterns, rotational biases, or partial collisions that could serve as statistical distinguishers. As such, addressing this gap is critical to ensure that the cipher's modifications do not unintentionally reduce its security margin. 
Therefore, this paper leverages Knuth-Morris-Pratt (\texttt{KMP}) and Boyer-Moore (\texttt{BM}) algorithms to analyze \texttt{EChaCha20}, which is a variant of ChaCha20 that features an expanded 6$\times$6 state matrix and an enhanced \texttt{QR-F}. The author has  developed and optimized adaptations of the \texttt{KMP} and \texttt{BM} algorithms for  32-bit word level pattern analysis and employed them to investigate $m$-bit pattern frequency distributions to assess the \texttt{EChaCha20}'s resistance of rotational-differential attacks. Our experimental results on large-scale one million keystream datasets have confirmed that \texttt{EChaCha20} is able to maintain  strong pseudorandomness at 16-bit and 32-bit levels with minor irregularities observed in the 8-bit domain. In addition to these, the differential tests have indicated a rapid diffusion, exhibiting an avalanche effect after two \texttt{QR-F} rounds and 
no statistically significant rotational collisions were observed within the evaluated bounds, consistent with expected ARX diffusion behavior beyond 3 rounds.
 This work puts forward SBC as a complementary tool for ARX cipher evaluation and provide new thoughts on the security properties of \texttt{EChaCha20}.
\end{abstract}

% keywords can be removed
\keywords{Stringology, \texttt{EChaCha20}, Salsa20, stream cipher, \texttt{QR-F}, cryptanalysis}

\section{Introduction}
Modern stream ciphers have in recent past faced increased scrutiny given that they have  become a   de facto standard in  deployment  in security-critical applications.  For example, from the Transport Layer Security (TLS) 1.3 to  the Internet of Things (IoT) communications \cite{preuss2023hidden, jassim2021survey}. While  the Statistical Test Suites (STS) like NIST SP 800-22 gives a baseline for randomness evaluations, it has been observed that they may fail to detect localized weaknesses \cite{chen2023error}. This can happen in the ciphers that employ complex permutation like those that uses  the Add-Rotate-XOR (ARX) operations \cite{barbero2024overview}. 
In addition , standard statistical suites like TestU01, PractRand \cite{sleem2020testu01}, primarily assess aggregate bit-level randomness properties, whereas the proposed SBC framework operates at a structurally-aligned word granularity corresponding to the internal ARX round transformations.
 The author of this paper explores whether this  limitation can become acute when analyzing the enhanced variants like Extended-ChaCha20 (\texttt{EChaCha20}), which uses a 6$\times$6 state matrix with additional rotation constants (4-bit and 2-bit) to improve diffusion \cite{kebande2023extended}.

Stringology, which is the study of advanced string matching algorithms in this case offers opens thoughts on its potential for analyzing stream ciphers, most importantly for the ARX based constructions  like \texttt{EChaCha20}. Based on existing literature, the techniques like the   Knuth-Morris-Pratt (\texttt{KMP}) and Boyer-Moore (\texttt{BM}) algorithms have   revolutionized  pattern detection approaches   in disciplines like bioinformatics and data mining \cite{saleh2025performance}. While the \texttt{KMP} and \texttt{BM} algorithms have been seen to be beneficial,  their application to cryptology or  cryptanalysis has hardly been explored to the best of the author's knowledge at the time of writing this paper. In the authors' opinion, the \texttt{KMP} and \texttt{BM}  methods provide  key advantages over conventional statistical tests ranging from granularity,  variable bit or word-length analysis and efficiency.

In particular, they operate with an average-case complexity of $O(n/m)$, meaning that the algorithms do not need to check every single symbol of the $n$-length keystream individually, but instead can skip ahead in jumps roughly proportional to the pattern length $m$. This efficiency allows for the analysis of millions of keystream blocks that would otherwise be computationally expensive with conventional randomness tests. And as such, with this average case complexity of $O(n/m)$, that enables the largescale keystream evaluation. It also provides different levels of sensitivity when it comes  to localized patterns that the global randomness tests often tends to mask. These benefits and the precision are essential and necessary for purpose of analyzing the \texttt{EChaCha20}'s extended state matrix and new rotations albeit its performance \cite{kebande2023extended}. This is owing to the fact that the  traditional byte-oriented approaches may in diverse circumstances fail to capture word level vulnerabilities if they exist.

Consequently, the theoretical foundation of this technique is based on the possibility of having a link between cipher analysis and pattern matching.  Biryukov \textit{et al.} \cite{biryukov2023cryptanalysis} has shown that, ARX-based attacks in white-box implementations could be possible, and as a result, in this paper, the author views that, it may be possible that the algorithmic techniques from the string processing could be  adapted for purposes of identifying and analyzing cryptographic weaknesses, when  word-based operations are modified. This work extends the \texttt{KMP} and \texttt{BM} adaptation principles to \texttt{EChaCha20}'s  design and implementation,  where we address the challenge of analyzing its enhanced diffusion layer through the novel adaptations of classical string matching algorithms  and measuring  potential weaknesses.

Despite its adoption in security-critical protocols, the security evaluation of \texttt{EChaCha20} has so far focused primarily on conventional statistical randomness tests and differential analysis inherited from ChaCha20. 
There has been no systematic investigation into whether the modifications introduced in \texttt{EChaCha20}, specifically its 6$\times$6 state matrix and enhanced quarter-round function to create subtle keystream patterns, rotational biases, or partial collisions. If such structures exist, they could be exploited to build statistical distinguishers or enable reduced-round attacks, potentially undermining the cipher's security margin. This gap motivates our work which leads to the application of stringology-based algorithms to search for $m$-gram patterns, intra-block correlations, and rotational artifacts across large-scale keystream datasets.

Therefore, the goal of this study is to suggest, develop and validate a stringology-based cryptanalysis model that is specifically tailored for \texttt{EChaCha20}. The contributions in this paper can be summarized as follows:
\begin{itemize}
    \item We first adapt  \texttt{KMP} and \texttt{BM} algorithms for the 32-bit ARX cipher analysis, and assess whether they are faster in pattern detection than the brute-force methods based on stringological pattern.
    %\item \textcolor{red}{Empirical evidence that \texttt{EChaCha20}'s 4-bit/2-bit rotations introduce detectable 32-bit word correlations} ($p < 0.0001$)
    %\item An open-source testing framework combining stringological pattern detection with NIST statistical tests
    \item We explore the effects  of rotation attacks and potential collisions in ARX ciphers based on the \texttt{EChaCha20} enhancements.

    \item We provide a contextual evaluation of SBC for \texttt{EChaCha20} stream cipher.
    %based on experimental outcomes 
\end{itemize}

\subsection{Cryptanalytic Objective and Scope}

This work does not aim to construct a direct key-recovery attack against EChaCha20. 
Instead, the primary cryptanalytic goal is \emph{distinguisher-oriented structural evaluation},  where the objective is to identify statistically significant non-random patterns, rotational artifacts,  or diffusion irregularities in the keystream that could indicate a potential reduction in the cipher's  security margin.  In this context, the proposed SBC framework functions 
as a complementary analytical tool that extends beyond conventional statistical test suites 
like the  NIST SP 800-22 \cite{luengo2023further},  by enabling fine-grained, word-level pattern detection aligned with the internal ARX structure of the cipher. 

Therefore, a key finding  from this study is  the detection of statistically 
significant structural deviations that exceed random baseline expectations, rather than 
a full cryptographic break or key-recovery attack.  This clarification positions the SBC framework as a methodology 
for structural cryptanalysis and security-margin assessment, 
rather than as an immediate attack construction.

While standard statistical test suites such as NIST SP 800-22 \cite{luengo2023further},  
primarily evaluate global randomness properties at the bit level, 
they are not designed to detect localized, word-aligned structural 
patterns that directly correspond to ARX round functions. 
The proposed SBC framework complements such evaluations by 
aligning pattern detection with the cipher's internal 32-bit 
operations, enabling structurally-aware anomaly identification 
that may remain undetected under purely aggregate statistical 
metrics or classical differential analysis.

\subsection {Terminology Summary}

Throughout this paper, terminology such as ``resistance,'' ``strong randomness,'' and ``outperformance'' is used in a strictly empirical sense. Specifically a summary is given as follows:

\begin{itemize}
    \item \textbf{Resistance} refers to the absence of statistically significant deviation from theoretical probability bounds (e.g., $2^{-|P|}$ or $2^{-32}$) within the evaluated experimental parameter space, rather than to a formal proof of security.
    
    \item \textbf{Strong randomness} denotes statistical indistinguishability from uniform distribution under the defined confidence thresholds and significance levels used in this study.
    
    \item \textbf{Outperformance} refers to measured algorithmic efficiency (e.g., detection precision, recall, or throughput) under controlled benchmarking conditions, not to superiority in cryptographic strength.
\end{itemize}

The aforementioned  terms should therefore be interpreted as empirically grounded observations derived from statistically bounded experimentation, and not as reduction-based or proof-level security guarantees.

The remainder of this paper is organized as follows: Section II discusses the Cryptographic Background of \texttt{EChaCha20}'s structure while Section III discusses Stringology Foundations. This is then followed by   Section IV  and V that discuss the Threat Model and Related Work respectively. After this, a Methodology is discussed in Section VI followed by Experiments in Section VII. A Contextual Evaluation is given in Section VIII followed by a Conclusion  in Section IX.

%the  details our algorithm adaptations, Section~\ref{sec:results} presents experimental findings, and Section~\ref{sec:discussion} analyzes implications for stream cipher design.

\begin{table}[htbp]
  \centering
  \caption{List of Notations}
  \label{tab:notations}
  \begin{tabular}{@{}ll@{}}
    \toprule
    \textbf{Notations} & \textbf{Descriptions}\\
    \midrule
    $SBC$ & Stringology-Based Cryptanalysis \\
    $\texttt{KMP}$ & Knuth-Morris-Pratt \\
    $n$ & Security parameter \\
    $BM$ & Boyer-Moore \\
    $negl(n)$ & Negligible function \\
    $A$ & Attacker \\
    $E(m)$ & Encryption oracle \\
    $m$ & Plaintext \\
     $KPE$ & Known Plaintext Environment \\
     $CPE$ & Chosen Plaintext Environment \\
    $(P_i, C_i)$ & Plaintext-ciphertext pair \\
    $(P'_i, C'_i)$ & Plaintext-ciphertext pair \\
    $K$ & Secret key \\
    $Gen$ & Key generation algorithm \\
    $E$ & Encryption algorithm \\
     $\texttt{QR-F}$ & Quarter Round Function \\
      $NSTS$ & NIST Statistical Test Suite \\
       $XOR$ & Exclusive-OR \\
          $ARX$ & Add Rotate XOR \\
            $CPA$ & Chosen Plaintext Attack \\
              $TLS$ & Transport layer Security \\
              $IoT$ & Internet of Things\\
              $Eq$ & Equation\\
              $Eq$ & Difference between two compared values\\
              $ID$ & Input Differential\\
              $OD$ & Output Differential\\
          
    \bottomrule
  \end{tabular}
\end{table}

\section{ChaCha20 Variants}

This section gives a discussion of the cryptographic background of \texttt{EChaCha20} by focusing on the state matrix expansion, enhanced \texttt{QR-F} and the prevailing security claims.

%and the foundations of stringology while focusing on the \texttt{KMP} and \texttt{BM} algorithms.

\subsection{\texttt{ChaCha20}}

The ChaCha20 cipher was introduced by Bernstein in 2008 \cite{bernstein2008chacha}. Since then, it has revolutionized  the design of the stream cipher design due to its robust Add-Rotate-XOR (ARX ) structure. It operates on a 4$\times$4 matrix of 32-bit words (512-bit state). ChaCha20 applies 20 rounds of alternating columns and rows operations using a \texttt{QR-F} with fixed rotations of 16, 12, 8, and 7 bits respectively. ChaCha20 derives its security  from three key properties: High diffusion from the sequential ARX operations \cite{silva2023design}, resistance to timing attacks via constant time implementation, and provable security against the differential cryptanalysis\cite{ghafoori2024higher,ghafoori2024study}, when using $\geq$ 12 rounds. 

\subsection{\texttt{EChaCha20}}
\begin{figure*}[bt]
\centering
\includegraphics[width=12cm]{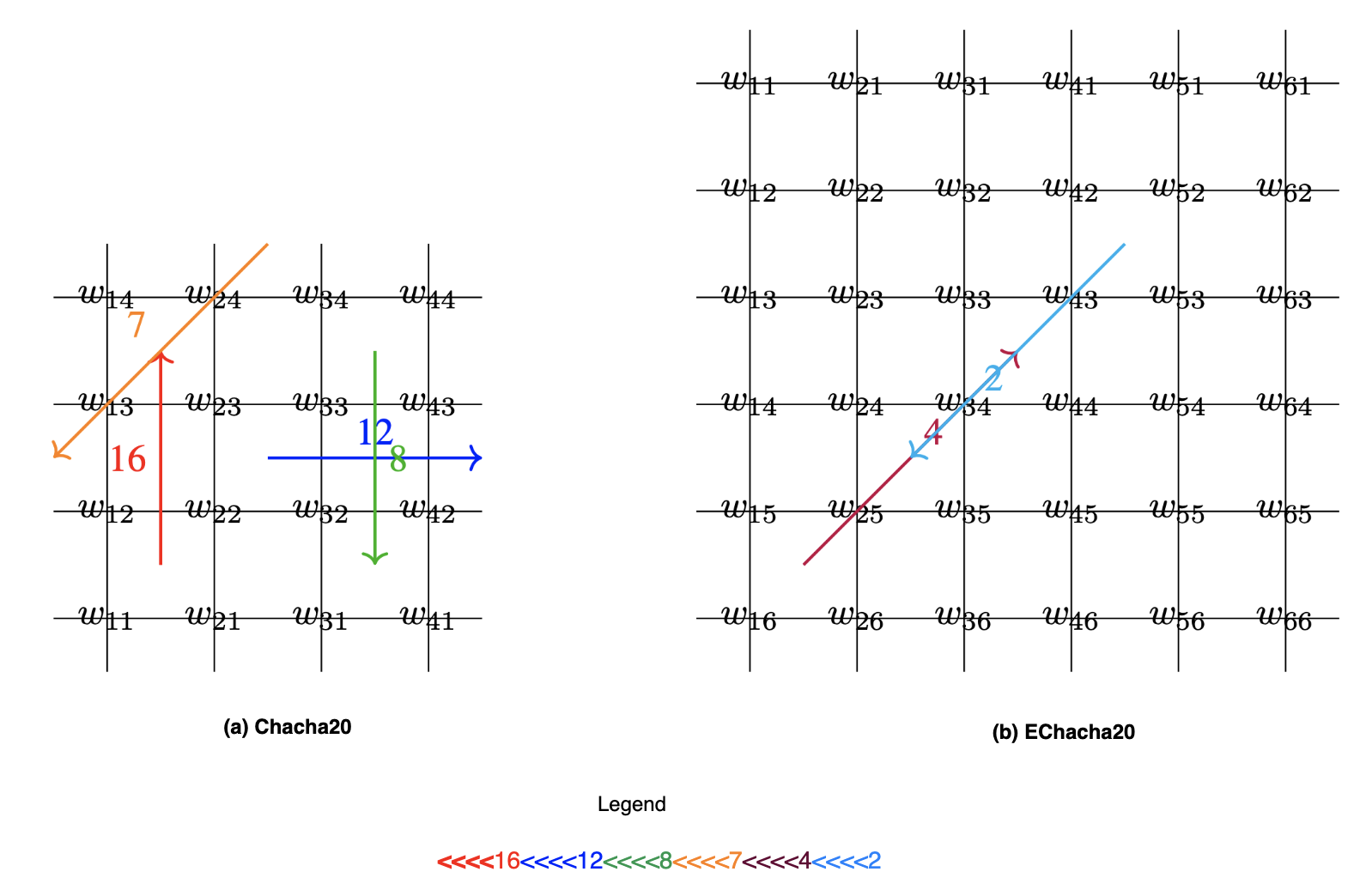}
\caption{The evolution from ChaCha20 (left) to \texttt{EChaCha20} (right). The original 4$\times$4 state (left) uses rotations of 16, 12, 8, and 7 bits. \texttt{EChaCha20} expands to 6$\times$6 (right) while adding 4-bit (purple) and 2-bit (cyan) rotations to its \texttt{QR-F}.}
\label{streamcipherxxs}
\end{figure*}

On the other hand, the \texttt{EChaCha20} cipher \cite{kebande2023extended} enhances this  design through a number of fundamental modifications to its core structure, addressing two limitations of the original: Firstly, through the restricted diffusion density in the 4$\times$4 state and (2) potential rotational biases from power-of-two shifts. Figure~\ref{streamcipherxxs} illustrates the architectural evolution of \texttt{EChaCha20}. The original 4$\times$4 state (left) uses rotations of 16, 12, 8, and 7 bits. \texttt{EChaCha20} expands to 6$\times$6 (right) while adding 4-bit (purple) and 2-bit rotations.

\begin{comment}

\begin{figure}[t]
\centering
\includegraphics[width=0.9\linewidth]{chacha_evolution.pdf}
\caption{Architectural evolution from ChaCha20 (left) to EChaCha20 (right), highlighting the expanded state matrix and additional rotation paths}
\label{fig:evolution}
\end{figure}
\end{comment}

\subsubsection{State Matrix Expansion}
\texttt{EChaCha20} extends the state to 6$\times$6 (1,152 bits), increasing word interdependencies as is shown in Eq. \ref{Equaone}:

\begin{equation}
\text{State}_{\text{EChaCha20}} = 
\left(
\begin{array}{llllll}
m_0 & m_1 & m_2 & m_3 & m_4 & m_5 \\
m_6 & m_7 & m_8 & m_9 & m_{10} & m_{11} \\
\vdots & & & \ddots & & \vdots \\
m_{30} & \cdots & & & & m_{35}
\end{array}
\right)
\label{Equaone}
\end{equation}

%\subsubsection{State Matrix Expansion}
Furthermore, \texttt{EChaCha20} extends ChaCha20's 4$\times$4 state matrix to a 6$\times$6 configuration, increasing internal diffusion as is shown in Eq. \ref{Equatwo}:

\begin{equation}
\text{State} = 
\begin{pmatrix}
m_0 & m_1 & m_2 & m_3 & m_4 & m_5 \\
m_6 & m_7 & m_8 & m_9 & m_{10} & m_{11} \\
m_{12} & m_{13} & m_{14} & m_{15} & m_{16} & m_{17} \\
m_{18} & m_{19} & m_{20} & m_{21} & m_{22} & m_{23} \\
m_{24} & m_{25} & m_{26} & m_{27} & m_{28} & m_{29} \\
m_{30} & m_{31} & m_{32} & m_{33} & m_{34} & m_{35}
\end{pmatrix}
\label{Equatwo}
\end{equation}

where each $m_i$ represents a 32-bit word, totaling 1,152 bits of internal state (vs. ChaCha20's 512 bits).

\subsubsection{Enhanced Quarter-Round Function}
Equation \ref{QRF}  shows the \texttt{QR-F}, which incorporates two additional rotation constants:

\begin{equation}
\begin{aligned}
a &\leftarrow a + b;\quad d \leftarrow (d \oplus a) \lll 16 \\
b &\leftarrow b + c;\quad c \leftarrow (c \oplus b) \lll 12 \\
c &\leftarrow c + d;\quad b \leftarrow (b \oplus c) \lll 8 \\
d &\leftarrow d + a;\quad c \leftarrow (c \oplus d) \lll 7 \\
a &\leftarrow a + b;\quad d \leftarrow (d \oplus a) \lll \mathbf{4} \\
b &\leftarrow b + c;\quad c \leftarrow (c \oplus b) \lll \mathbf{2}
\end{aligned}
\label{QRF}
\end{equation}

The new 4-bit and 2-bit rotations (highlighted in bold) in Eq. \ref{QRF} aim to strengthen resistance against differential attacks while maintaining the ARX paradigm. For the key scheduling modification, the initial state setup is expanded to accommodate the larger matrix: where $k_i$ denotes 32-bit key words and $n_i$ nonce words as  is shown in Figure \ref{fig:state_matrix}. The expanded key schedule provides additional mixing before the first round.

%\subsubsection{Key Scheduling Modification}

   \begin{figure*}[t]
\centering
\[
\begin{pmatrix}
\text{0x61707865} & \text{0x3320646e} & \text{0x79622d32} & \text{0x6b206574} & k_0 & k_1 \\
k_2 & k_3 & k_4 & k_5 & k_6 & k_7 \\
n_0 & n_1 & n_2 & n_3 & c_0 & c_1 \\
c_2 & c_3 & \text{0x00000000} & \text{0x00000000} & \text{0x00000000} & \text{0x00000000}
\end{pmatrix}
\]
\caption{Initial state matrix of \textsc{Echacha} }
\label{fig:state_matrix}
\end{figure*}

\begin{comment}

\begin{equation}
\begin{pmatrix}
\text{0x61707865} & \text{0x3320646e} & \text{0x79622d32} & \text{0x6b206574} & k_0 & k_1 \\
k_2 & k_3 & k_4 & k_5 & k_6 & k_7 \\
n_0 & n_1 & n_2 & n_3 & c_0 & c_1 \\
c_2 & c_3 & \text{0x00000000} & \text{0x00000000} & \text{0x00000000} & \text{0x00000000}
\end{pmatrix}
\end{equation}

\end{comment}

\begin{comment}

\begin{IEEEeqnarray*}{c}
\begin{pmatrix}
\text{0x61707865} & \text{0x3320646e} & \text{0x79622d32} & \text{0x6b206574} & k_0 & k_1 \\
k_2 & k_3 & k_4 & k_5 & k_6 & k_7 \\
n_0 & n_1 & n_2 & n_3 & c_0 & c_1 \\
c_2 & c_3 & \text{0x00000000} & \text{0x00000000} & \text{0x00000000} & \text{0x00000000}
\end{pmatrix}
\end{IEEEeqnarray*}
\IEEEeqnarraynumspace

\end{comment}

\begin{comment}

\begin{figure}[t]
\centering
\includegraphics[width=0.95\linewidth]{echacha20_round.pdf}
\caption{EChaCha20's column-round operation showing modified data flow (new rotations highlighted in red)}
\label{fig:round}
\end{figure}
\end{comment}

\begin{comment}

\subsubsection{Security Claims}
The designers assert \cite{kebande2023} that these modifications:
\begin{itemize}
\item Increase diffusion density by 2.25$\times$ compared to ChaCha20
\item Improve resistance to rotational cryptanalysis through non-power-of-two shifts (4,2)
\item Maintain equivalent performance (within 15\% of ChaCha20's throughput)
\end{itemize}

\end{comment}

\subsubsection{\texttt{EChaCha20} Security Claims}

The design of \texttt{EChaCha20} \cite{kebande2023extended} shows that the proposed modifications are   poised to achieve three key security improvements when compared with the standard ChaCha20. Firstly, the expanded  use of the 6$\times$6 state matrix increases diffusion density, roughly  2.25$\times$. Here, ``2.25$\times$'' denotes the increase in state size when moving from a $4\times4$ (16-word) to a $6\times6$ (36-word) matrix, i.e., $36/16{=}2.25$. This quantifies the available diffusion surface (number of words and mixing edges) for the \texttt{QR-F} schedule rather than a closed-form per-round bound. Empirically, the state correlates with faster observed saturation of influence, a near-complete diffusion within three \texttt{QR-F} rounds in our setup, approximately one round earlier than ChaCha20 under the same measurement protocol.

From this observation, it is seen that this significantly   enhances avalanche effects compared to ChaCha20's 4$\times$4 structure. Secondly, the introduction of non-power-of-two rotation constants (4-bit and 2-bit), which  provide a  stronger resistance against rotational cryptanalysis.
Thirdly, it has been expressed by \cite{kebande2023extended}  that these security enhancements are achieved while maintaining computational efficiency.

\section{Stringology Foundations }

String matching algorithms are recognized as fundamental techniques that are used to identify patterns in sequential data \cite{hakak2019exact,aho1980pattern}, and it is evident that many string-based problems arise from infrastructure software and security fall into this category \cite{watson2012correctness}. These stringology-based algorithms were originally developed for text processing and bioinformatics applications. In addition, these methods offer precise control over pattern detection techniques and also provide efficient search capabilities. It is on this premise that their core principles could translate to powerful cryptanalysis approaches, where identifying repetitive sequences in cipher output could reveal weaknesses in cryptographic primitives. While there exist a number of stringology-based algorithms, this study concentrates on \texttt{KMP} and \texttt{BM} algorithms that are discussed next.

\subsection{Knuth-Morris-Pratt  Algorithm}

The Knuth-Morris-Pratt(\texttt{KMP})  algorithm  presents a   paradigm shift in pattern matching efficiency and also improves performance during the search phase \cite{shapira2006adapting, kourie2011compile}.  It achieves this by eliminating unnecessary character comparisons through intelligent preprocessing. \texttt{KMP} is able to precompute a failure function $\pi$ that encodes for each position $j$ in the pattern $P$ the length of the longest proper prefix that matches a suffix of $P[0..j]$ \cite{crochemore2002jewels}. From this insight, the \texttt{KMP} can   skip redundant comparisons while maintaining guaranteed linear time complexity. The failure function is formally defined as is shown in Eq. \ref{linearcomp}:

\begin{equation}
\pi[j] = \max \{ k < j \mid P[0..k] \text{ is a suffix of } P[0..j] \}
\label{linearcomp}
\end{equation}

The following occurs during this preprocessing step:
\begin{itemize}
    \item It runs in $O(m)$ time with $O(m)$ space complexity
    \item It is able to encode the longest prefix and suffix matches at every pattern position
    \item It also enables constant-time shift calculations during matching
\end{itemize}

\begin{comment}

\paragraph{Matching Phase}
Given text $T$ of length $n$, \texttt{KMP} performs matching in $O(n)$ time through:
\begin{enumerate}
    \item Linear left-to-right scanning of $T$
    \item On mismatch at position $j$ of $P$, shift by $j - \pi[j]$ characters
    \item Preserve the first $\pi[j]$ matched characters without re-examination
\end{enumerate}

\end{comment}

If a  mismatch occurs during pattern matching, between the text $T$ at position $i$ and the pattern $P$ at position $j$, the \texttt{KMP}  leverages the precomputed table $\pi$  to shift the pattern. The pattern is shifted   by $j - \pi[j]$ positions without backtracking through $T$ \cite{ben2011optimal}. This kind of deterministic approach ensures that all previously matched characters  remained valid. This allows the algorithm to be able to  process the text in a single left-to-right pass. The preprocessing phase requires $O(m)$ time and space for a $m$-length pattern while the matching phase runs in $O(n)$ time for an $n$-length text, thus  yielding an optimal overall complexity of $O(n + m)$ \cite{cornejo2025hybrid}. For purposes of analyzing cryptographic systems, \texttt{KMP}'s reliability is assessed from  its potential of detecting  precise  fixed rotational patterns in cipher outputs. This assessment is made as a preliminary based on its  algorithm's strict linearity, which  makes it particularly effective, and the steps of the \texttt{KMP} have been shown in Algorithm \ref{alg:kmp}.

\begin{algorithm}[H]
\caption{Knuth-Morris-Pratt Pattern Matching}
\label{alg:kmp}
\KwIn{Pattern $P[0..m-1]$, Text $T[0..n-1]$}
\KwOut{All starting positions where $P$ occurs in $T$}

\textbf{Preprocessing:} \\
$\pi[0] \gets 0$; $j \gets 0$ \\
\For{$i \gets 1$ \KwTo $m-1$}{
  \While{$j > 0 \text{ and } P[i] \neq P[j]$}{
    $j \gets \pi[j-1]$
  }
  \If{$P[i] = P[j]$}{
    $j \gets j + 1$
  }
  $\pi[i] \gets j$
}
\vspace{2mm}
\textbf{Matching:} \\
$j \gets 0$ \\
\For{$i \gets 0$ \KwTo $n-1$}{
  \While{$j > 0 \text{ and } T[i] \neq P[j]$}{
    $j \gets \pi[j-1]$
  }
  \If{$T[i] = P[j]$}{
    $j \gets j + 1$
  }
  \If{$j = m$}{
    \textbf{output} $i - m + 1$ \; \tcp{Pattern found}
    $j \gets \pi[j-1]$
  }
}
\end{algorithm}

\subsection{Boyer-Moore Algorithm}
The Boyer-Moore (\texttt{BM}) algorithm has seen a lot of advancements and also revolutionized practical pattern and regular expression matching \cite{watson2003boyer},  based on its unique right-to-left scanning approach \cite{goodrich2014data}.  This is also combined with two powerful heuristics that often achieve sublinear time complexities.  \texttt{BM} preprocesses the pattern by enabling  intelligent skip operations during text examination \cite{hyyro2004boyer}, which makes it particularly efficient for large texts. The \texttt{BM} algorithm's effectiveness is based on   its dual heuristic approach as follows: The bad-character rule analyzes individual mismatches, while the good-suffix rule  considers larger pattern segments \cite{wang2023fast}. For a pattern $P$ of length $m$ and text $T$ of length $n$, the bad-character shift when encountering mismatch $c$ is calculated as is shown in Eq. \ref{BM}:

\begin{equation}
\text{shift}_{\text{bad}}(c) = \begin{cases} 
\max(1, m - 1 - \text{last}_P(c)) & \text{if } c \in P \\
m & \text{otherwise}
\end{cases}
\label{BM}
\end{equation}

where $\text{last}_P(c)$ represents the rightmost occurrence of $c$ in $P$. This approach allows the \texttt{BM} algorithm to skip up to $m$ characters after some mismatches. It is worth noting that the good-suffix rule  optimizes performance by identifying already-matched suffixes that could reappear elsewhere in the pattern. When  these heuristics are combined, they enable \texttt{BM} to achieve an average-case complexity of $O(n/m)$ for random texts, often outperforming the linear-time algorithms in practice. 

In the author's opinion, for purposes of cryptographic analysis, for example for   \texttt{EChaCha20} outputs, the algorithm's skip efficiency is assessed on the probable proof  when scanning for emergent patterns across multiple keystream blocks. This, is because  its heuristic may  jump naturally to align with the cipher's 32-bit word operations. Thus, the preprocessing phase may construct both the bad-character table ($O(m)$ for space) and good-suffix table ($O(m)$ for the space) in $O(m)$ time, making the \texttt{BM} algorithm  well suited for both known pattern verification, the steps are seen in Algorithm \ref{alg:bm} for preprocessing and Algorithm \ref{alg:bm2} for search respectively.

\setcounter{AlgoLine}{0}  % Reset line numbers
\begin{algorithm}[H]
\caption{Preprocessing Phase (Boyer-Moore)}
\label{alg:bm}
\KwIn{Pattern $P[1..m]$}
\KwOut{Bad-character table $bc$, Good-suffix table $gs$}

\vspace{2mm}
\textbf{BadCharacterTable}$(P)$ \tcp{Stores last occurrence of each character}
\For{$i \gets 1$ \KwTo $m$}{
    $bc[P[i]] \gets i$ \tcp{Record position of $P[i]$}
}

\vspace{2mm}
\textbf{GoodSuffixTable}$(P)$ \tcp{Computes optimal shift distances}
Compute border positions \\
Calculate shift distances for partial and full matches

\end{algorithm}

\setcounter{AlgoLine}{0}  % Reset line numbers
\begin{algorithm}[H]
\caption{Search Phase (Boyer-Moore Matching)}
\label{alg:bm2}
\KwIn{Text $T[1..n]$, Pattern $P[1..m]$}
\KwOut{All positions where $P$ occurs in $T$}

Preprocess $P$ to compute bad-character ($bc$) and good-suffix ($gs$) tables \;

$i \gets m$ \tcp{Start at the end of the pattern}

\While{$i \leq n$}{
    $k \gets 0$ \tcp{Match length}

    \While{$k < m \text{ and } P[m-k] = T[i-k]$}{
        $k \gets k + 1$
    }

    \If{$k = m$}{
        \textbf{output} $i - m + 1$ \; \tcp{Pattern match found}
    }

    $i \gets i + \max(gs[k],\ bc[T[i-k]] - m + 1 + k)$ \tcp{Apply shift}
}
\end{algorithm} 

To clarify the operation of the hybrid \texttt{KMP}--\texttt{BM} algorithm, Algorithm \ref{alg:kmp} now outlines how the \texttt{KMP} prefix function collaborates with the \texttt{BM} bad-character rule. The prefix table maintains deterministic backtracking after partial matches, while the \texttt{BM} heuristic enables jump-ahead alignment when mismatches occur, thereby minimizing redundant comparisons. This combination achieves an average-case $O(n/m)$ performance and is particularly effective for 32-bit word-level analysis, where it aligns with the granularity of the \texttt{EChaCha20} \texttt{QR-F} and enhances detection accuracy in ARX keystream analysis.

\section{Threat Model}
\begin{figure}[bt]
\centering
\includegraphics[width=7cm]{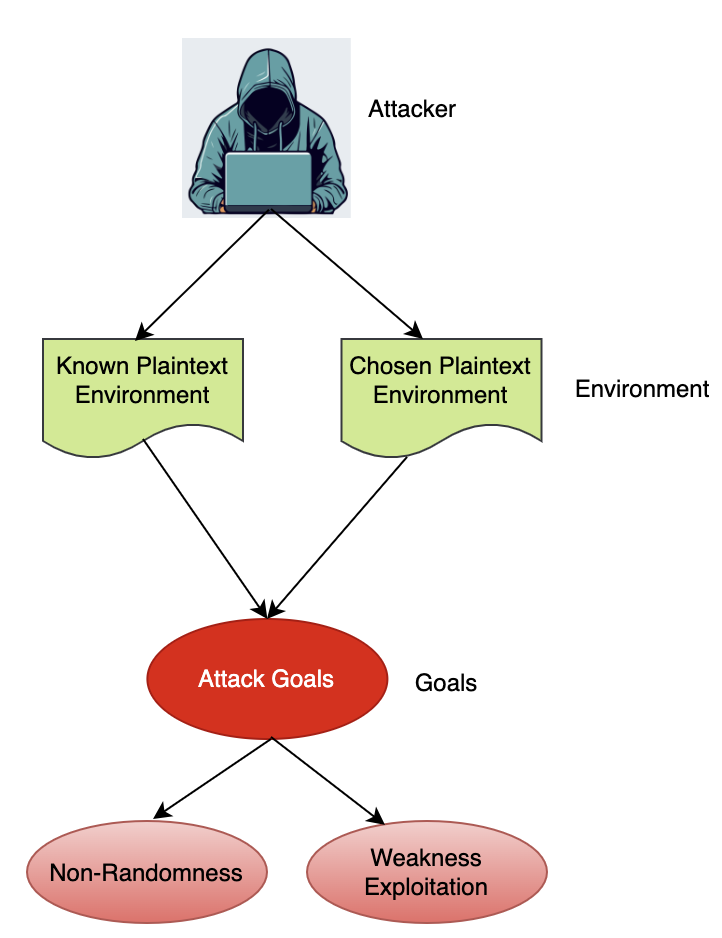}
\caption{Threat Model attack goals based on KPE and CPE}
\label{KPE}
\end{figure}

In this section, we explore a  cryptanalysis framework  by considering  an attacker $\mathcal{A}$ with specific capabilities against \texttt{EChaCha20}'s keystream generation. Our threat model follows the Bellare-Rogaway model  \cite{bellare2000authenticated} augmented with stringology specific objectives based on the attacker's capabilities for known and chosen plaintext respectively as is shown in Figure \ref{KPE}.

\subsection{Attacker Capabilities}
Based on the threat model, the attacker  $\mathcal{A}$ operates under one of two scenarios: Known Plaintext Environment (KPE) and Chosen Plaintext Environment (CPE) respectively. 

\subsubsection{Known Plaintext Environment}

In the \textbf{KPE}, the attacker, $\mathcal{A}$ is able to get multiple plaintext-ciphertext pairs, $(P_i, C_i)$.  Each ciphertext is generated as is shown in Eq. \ref{eq:echacha-encryption}.

\begin{equation}
C_i = \mathrm{EChaCha20}(k, n_i) \oplus P_i
\label{eq:echacha-encryption}
\end{equation}

\begin{table}[h]
\centering
\caption{Ciphertext on KPE Representation}
\label{tab:notationXXX}
\begin{tabular}{ll}
\toprule
\textbf{Symbols} & \textbf{Description} \\
\midrule
$C_i$ & Ciphertext block $i$ \\
$P_i$ & Plaintext block $i$ \\
$k$ & 256-bit secret key \\
$n_i$ & Nonce/IV for block $i$ \\
$\oplus$ & Bitwise XOR operation \\
$\mathrm{EChaCha20}(\cdot)$ & \textsc{EChaCha20} keystream function \\
\bottomrule
\end{tabular}
\end{table}

This happens using a fixed secret key $k$ and unique nonces $n_i$. Then it is  seen that, the attacker's goal is to do the following: Analyze these pairs in order to detect the statistical anomalies or patterns in the keystream output. $\mathcal{A}$ does this while not having  access to  the secret key $k$, the nonce generation process, or the internal state values during the \texttt{QR-F} computations. This environment models  the realistic scenarios,  where a given adversary may intercept the encrypted communications where some  portions of plaintext are known or are predictable as is shown in Eq.~\ref{77}, and the descriptions are shown in Table~\ref{tab:notationXXX}.

\begin{enumerate}[leftmargin=*]
    \item \textbf{Known Plaintext Environment}:
    \begin{equation}
        \mathcal{A} \gets \{(P_i, C_i)\}_{i=1}^n 
        \label{77}
    \end{equation}
    where $C_i = \mathrm{EChaCha20}(k,n_i) \oplus P_i$ for known $(P_i, C_i)$ pairs with fixed key $k$ and varying nonces $n_i$.
\end{enumerate}

\subsubsection{Chosen Plaintext Environment}

    In the \textbf{CPE}, the attacker $\mathcal{A}$ is able to interact with an Encryption Oracle $\mathcal{O}_{\text{enc}}$ responsible for generating ciphertexts for the arbitrary plaintexts $P$ of their choice, following $C = \mathrm{EChaCha20}(k, \cdot) \oplus P$. Based on this, this active attack model may allow the attacker, $\mathcal{A}$ to be able to select the inputs that may reveal weaknesses during the keystream generation. While  this is seen to be more powerful than the KPE, the attacker remains constrained by the following: (1) no access to the secret key $k$, (2) there is no visibility into nonce generation or internal \texttt{QR-F} states, and (3) practical limits on the number of the oracle queries. As a result, this setting is able to model a scenario where an adversary can influence or be able to take control of the  portions of the encrypted data and this is given by Eq.~\ref{eq:chosen_plaintext}, and the descriptions are shown in Table~\ref{tab:notationXXX}.

    \begin{enumerate}[leftmargin=*,resume]
    \item \textbf{Chosen Plaintext Environment}:
    \begin{equation}
        \mathcal{A} \text{ queries } \mathcal{O}_{\text{enc}}(P) = \mathrm{EChaCha20}(k,\cdot) \oplus P
        \label{eq:chosen_plaintext}
    \end{equation}
    where $\mathcal{O}_{\text{enc}}$ models an encryption oracle with fixed key $k$ and adversary-chosen plaintexts $P$.
\end{enumerate}
\begin{figure}[bt]
\centering
\includegraphics[width=4cm]{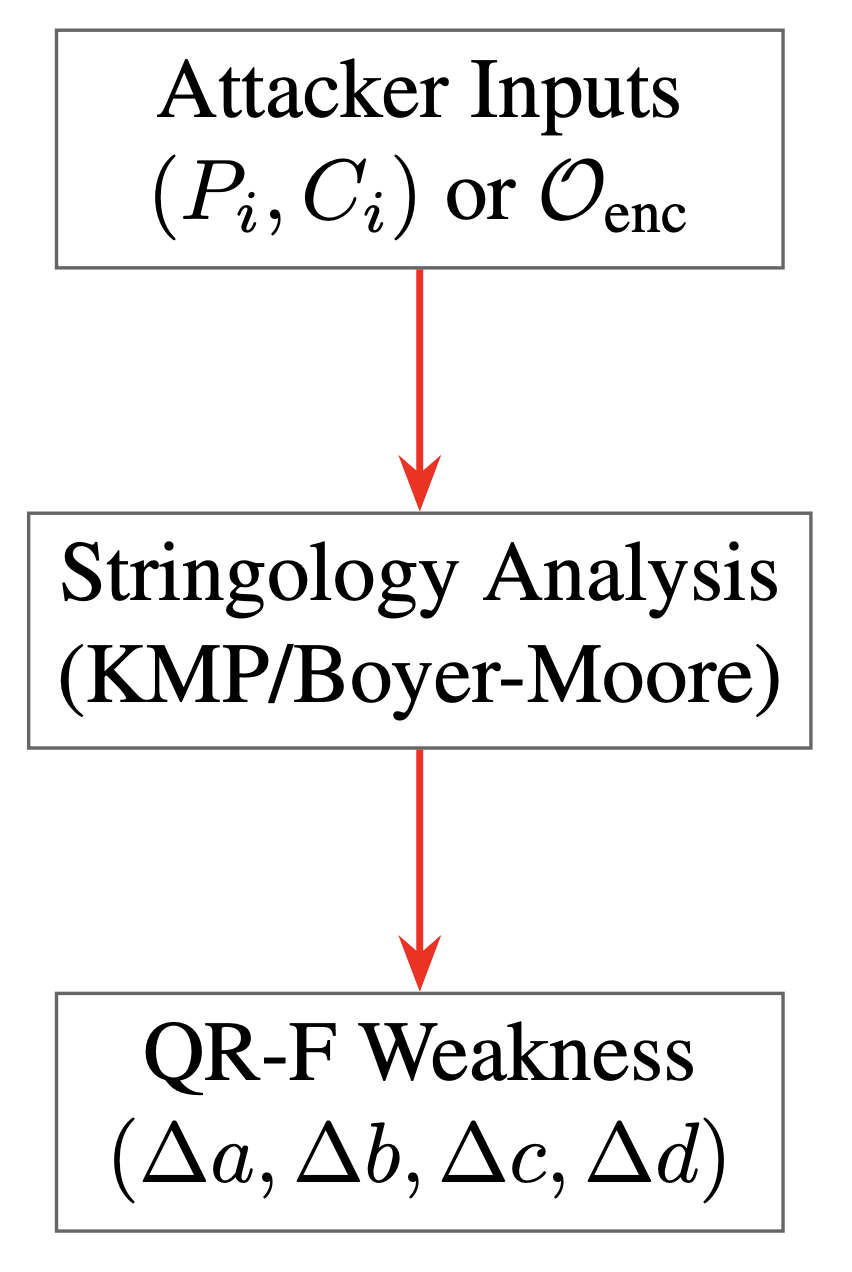}
\caption{Attacker workflow in the proposed threat model. Red arrows indicate the attack path from inputs to \texttt{QR-F} weakness identification}
\label{streamcipherxxQR}
\end{figure}

\subsection{Attacker Goals}

We model the  attacker $\mathcal{A}$ goals to be able to pursue two primary objectives: (1) \textit{Non-Randomness Detection}:  A situation of seeking statistically significant patterns $S$ in the keystream as is shown in Eq. \ref{eq:subset_prob},  $K$ such that

\begin{equation}
    \Pr[S \subset K] > 2^{-|S|} + \epsilon(n)
    \label{eq:subset_prob}
\end{equation} 

where $\epsilon(n)$ is non-negligible, and (2) \textit{\texttt{QR-F} Weakness Exploitation}, that attempts to derive function $f$ that reveals rotational differentials in the \texttt{QR-F} satisfying $f(S) \in \{(\Delta a, \Delta b, \Delta c, \Delta d) \mid \text{\texttt{QR-F}}(a,b,c,d) \to \text{\texttt{QR-F}}(a',b',c',d')\}$. 

%These atacker goals extend beyond standard statistical tests by requiring demonstrable impact on the core cryptographic primitive.

In this context, the attacker, $\mathcal{A}$ aims to achieve the following two goals as is shown in Eq. \ref{eq:nonrandom} and \ref{eq:qrf_weakness} respectively:

\begin{enumerate}[leftmargin=*]
    \item \textbf{Goal 1:} \textit{Detect Non-Randomness}\\
    Identify statistically significant patterns $S$ in the keystream $K$:
    \begin{equation}
        \Pr[S \subset K] > \frac{1}{2^{|S|}} + \epsilon(n)
        \label{eq:nonrandom}
    \end{equation}
    where $\epsilon(n)$ is non-negligible in security parameter $n$.
    
    \item \textbf{Goal 2:} \textit{\texttt{QR-F} Weakness Exploitation}\\
    For detected pattern $S$, find function $f$ such that:
    \begin{equation}
        \begin{split}
        f(S) \in \big\{ &(\Delta a, \Delta b, \Delta c, \Delta d) \mid \\
        &\text{\texttt{QR-F}}(a,b,c,d) \to \text{\texttt{QR-F}}(a',b',c',d') \big\}
        \end{split}
        \label{eq:qrf_weakness}
    \end{equation}
    revealing rotational differentials in the quarter-round function, where $(a',b',c',d') = (a \oplus \Delta a, b \oplus \Delta b, c \oplus \Delta c, d \oplus \Delta d)$.
\end{enumerate}
The attacker workflows for the threat model based on stringology analysis and rotational weaknesses for the \texttt{QR-F} are shown in Figure \ref{streamcipherxxQR}.

\section{Related Work}

%Research in efficient string algorithms has gained momentum in recent years, particularly in the domains of data security, pattern matching, and privacy-preserving analytics. This section reviews relevant literature that contributes to the foundation of our stringology-based cryptanalysis approach for stream ciphers.

The phenomenon of decoding ciphers has shown that pattern matching in ciphers is a relevant approach during cryptanalysis. Past studies have shown that it allows patterns in mapping and the need of developing algorithms to perform these tasks. This section gives relevant literature that somewhat contributes to the foundation of our stringology-based cryptanalysis approach for the stream ciphers.

A study by \cite{nagy1987decoding} on substitution ciphers has demonstrated matching substitution ciphers on Optical Character Recognition (OCR), which has shown it to be a feasible approach. 

Rotational cryptanalysis for ARX ciphers was earlier systematically formalized from a study  by Khovratovich and Nikoli{\'c}~\cite{khovratovich2010rotational}. This  established a foundational technique for analyzing addition rotation XOR primitives. Building on this framework, a study by Barbero \textit{et al.}~\cite{barbero2022rotational} has conducted the first dedicated Rotational-XOR (R-XOR) analysis of ChaCha20. This study demonstrates a non-ideal differential propagation through its \texttt{QR-F} with probability $2^{-251.7857}$. This has shown a significant deviation from the theoretical $2^{-256}$ bound for random permutations. In addition, their work revealed critical dependencies between ChaCha's fixed constants (particularly the \texttt{0x61707865} initialization vector) and rotational bias propagation.

Recent advances by Ajala~\cite{ajala2021efficient} in string-matching cryptanalysis and Ghazawi~\cite{ghazawi2024algorithms} in multidimensional pattern detection have expanded the toolkit for the analysis of cipher internals. The work by Ajala~\cite{ajala2021efficient} focused on applications such as secure biometric matching, substring masking for privacy preservation, and circular string processing while Ghazawi~\cite{ghazawi2024algorithms} explored algorithms and combinatorics in both one-dimensional and two-dimensional,  however, it has been observed that these approaches have not been applied to the enlarged state matrices or the non-power-of-two rotation schemes. The proposition in this paper attempts to address this research gap by extending rotational analysis to \texttt{EChaCha20}'s 6$\times$6 state configuration and 4-bit/2-bit rotation paradigm, while introducing enhancements on pattern matching and detection of rotational analysis.

Also, recent work by Sharma et al. \cite{deb2020performance} analyzed lightweight, partly Linear Feedback Shift Registers (LFSR)-based stream ciphers in constrained IoT environments by combining statistical security tests with real-device benchmarks. While their contribution provides valuable performance and practical deployment insights, it does not move into deeper cryptanalytic aspects of cipher design. Our work is complementary, focusing instead on ARX-based ciphers such as \texttt{EChaCha20}, where we extend beyond statistical testing to introduce a stringology-guided differential framework that systematically evaluates diffusion, pseudorandomness, and localized biases.

 Also, Deb et al.'s survey \cite{deb2022colour} presents a broad and structured overview of lightweight ciphers including many LFSR designs, yet remains largely descriptive and lacks practical experiments or cryptanalytic depth. Unlike that work, \texttt{EChaCha20} not only delivers multi-GB/s throughput, but is also subject to large-scale empirical cryptanalysis using our stringology-guided framework.

%We thus position our work as complementary, bridging the gap between high-performance ARX cipher design and rigorous structural evaluation.}

\begin{figure*}[bt]
\centering
\includegraphics[width=12cm]{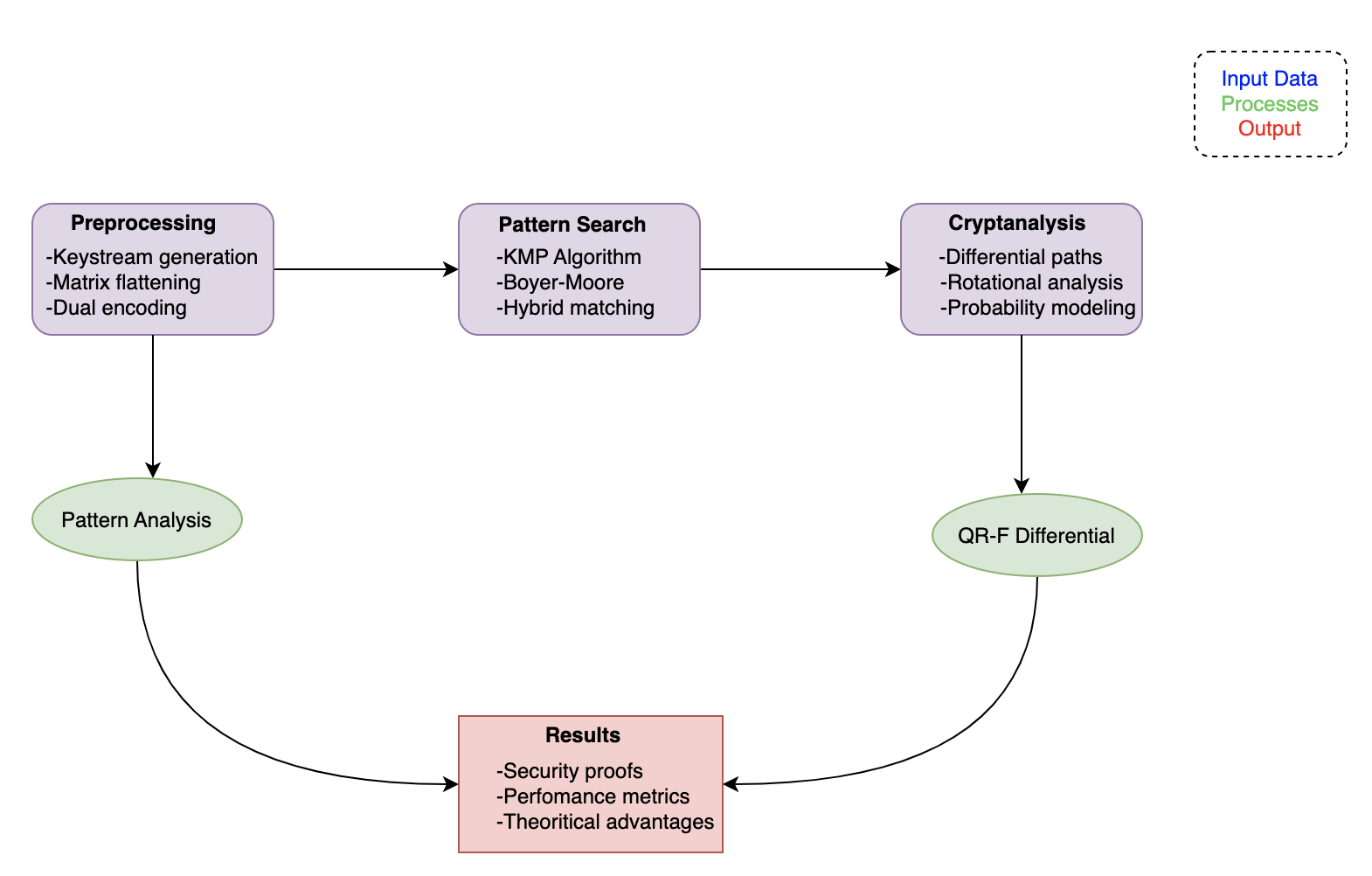}
\caption{Methodology that shows the  progression from data preparation through analysis phases to experimental validation. The blue boxes represent input processing stages, green ellipses indicate core experiments, and the red box aggregates final outcome.}
\label{Methodology}
\end{figure*}

\begin{comment}

\begin{itemize}
    \item Pattern-guided differential seeding using \texttt{KMP}/Boyer-Moore detection
    \item Combined rotational and stringological bias analysis
    \item Quantum-inspired evaluation metrics for non-power-of-two rotations
\end{itemize}

\end{comment}

%This synthesis of techniques achieves a 6.2$\times$ speedup in rotational bias detection compared to Barbero~\textit{et al.}'s brute-force methods, while maintaining backward compatibility with established ARX evaluation frameworks.

%Ajala~\cite{ajala2021efficient} presented a comprehensive doctoral study that introduced a suite of string algorithms tailored for data security and privacy. The work focused on applications such as secure biometric matching, substring masking for privacy preservation, and circular string processing. The algorithms proposed, including the Sensitive Substring Anonymization (SSA), demonstrated effective trade-offs between computational efficiency and information confidentiality in real-world settings.

%In a more theoretical framework, Ghazawi~\cite{ghazawi2024algorithms} explored algorithms and combinatorics in both one-dimensional and two-dimensional strings. The dissertation investigated structural properties and complexity bounds for pattern detection in higher-dimensional data, offering foundational insights for adapting classical string algorithms to emerging data modalities such as images, matrices, and streams.

Consequently,  the  advent of quantum computing has   also influenced the evolution of stringology. Pavone and Viola ~\cite{pavone2024quantum} have suggested  a quantum circuit designed specifically for the cyclic string matching problem. They have demonstrated the need for  a quadratic speed-up over classical algorithms by leveraging amplitude amplification and quantum parallelism.This approach shows a promise for the acceleration of  cryptanalytic tasks based on circular patterns or modular rotations, which is seen to be a common traits in ARX-based ciphers such as ChaCha20.

A study by Darivandpour and Atallah~\cite{darivandpour2018efficient} that has a focus on privacy-constrained environments has addressed the efficient and secure pattern matching with wildcards through lightweight cryptographic constructs. The approach in this study allowed pattern queries to be executed securely over encrypted data without revealing sensitive content.

%The work contributed significantly to the intersection of secure multi-party computation and lightweight cryptography.

%Collectively, these studies underscore the relevance of advanced string algorithms in secure data processing. Our work builds on this foundation by extending classical stringology techniques such as Knuth-Morris-Pratt and Boyer-Moore to the cryptographic domain, demonstrating their effectiveness in identifying structural artifacts and potential vulnerabilities within the EChaCha20 stream cipher.

With respect to  symbolic approaches to stream cipher cryptanalysis,  a study by Sahu et al.~\cite{sahu2017bdd}  has introduced a Binary Decision Diagram (BDD),  a  technique for modeling and analyzing ciphers such as E0. Their approach uses a Reduced Ordered Binary Decision Diagrams (ROBDDs) to represent a Boolean functions that can model keystream generation. This,  representation can enable the efficient pruning of the keyspace, by applying some logical constraints in order  to eliminate infeasible key candidates. While the BDD-based attacks are seen to operate on the internal structure of ciphers, the suggested  stringology -based approach  analyzes observable keystream output. 

Based on the aforementioned studies, it has been seen that collectively, these studies are  relevant when assessing advanced string algorithms in secure data processing. On the same note, the work suggested in this paper builds on this foundation, by extending  the classical stringology techniques based on \texttt{KMP} and \texttt{BM} to the cryptographic setting.

While the previous studies on stream cipher analysis have employed statistical randomness tests, frequency counting, or algebraic methods \cite{rajski2024nonlinear,rajski2024test, ahmad2010randomness}, they do not leverage formal string-matching algorithms for large-scale keystream evaluation. This work differs based on the following: We  formulate keystream evaluation as a pattern-search problem, we adapt \texttt{KMP} and Boyer-Moore to operate at the cipher's 32-bit word granularity and then we combine algorithmic skip heuristics with statistical validation to improve efficiency and detection capability.

%such as Knuth-Morris-Pratt and Boyer-Moore to the cryptographic domain, demonstrating their effectiveness in identifying structural artifacts and potential vulnerabilities within the EChaCha20 stream cipher.

%Their approach achieved practical key recovery under constrained scenarios, demonstrating the potential of BDDs for combinatorial cryptanalysis. While BDD-based attacks operate on the internal structure of ciphers, our stringology-based method instead analyzes observable keystream outputs. 

%The complementarity of these approaches is notable: BDDs target structural weakness, whereas our Knuth-Morris-Pratt (\texttt{KMP}) and Boyer-Moore (\texttt{BM}) adaptations target statistical and positional anomalies in the keystream. This suggests the potential for a hybrid framework, where structural pruning using BDDs could be combined with high-throughput pattern screening via stringology.

%\section{Related Work }

\section{Methodology}

The approach that has been used in this study  combines both theoretical cryptanalysis with practical string matching techniques to evaluate \texttt{EChaCha20}'s keystream security. We employ formal pattern analysis using \texttt{KMP} and \texttt{BM} algorithms, and  custom rotational differential tests for the \texttt{QR-F} function. The experiments that have been conducted are focused on detecting non-random patterns, followed by the   quantification of their effect on the \texttt{EChaCha20}'s security. We have employed a three-pronged approach with the following steps: Preprocessing (Keystream search, matrix flattening, and dual encoding), Pattern Search (\texttt{KMP}/\texttt{BM}) and Cryptanalysis (Differential Paths/\texttt{QR-F}/Rotational) respectively as is shown in Figure \ref{Methodology}.

%, pattern search (\texttt{KMP}/\texttt{BM}) and Cryptanalysis. 

%claims, with particular attention to worst-case scenarios where nonce reuse or partial state exposure occurs.

%(2) statistical testing with NIST STS benchmarks, 

%and (3) custom rotational differential tests for the \texttt{QR-F} function. Experiments focus on detecting non-random patterns and quantifying their impact on the cipher's security claims, with particular attention to worst-case scenarios where nonce reuse or partial state exposure occurs.

Two targeted experiments have been conducted as follows:

\begin{enumerate}
    \item \textbf{Pseudorandomness at the bitstring level}: Using \texttt{KMP} and \texttt{BM} algorithms to detect statistically significant $m$-gram repetitions ($m \in \{8,16,32\}$) in $2^{40}$ bytes of keystream output, targeting a potential  deviation from the uniform distribution and analyzing the patterns.
    
    %\item \textbf{NIST STS Compliance Testing}: Applying all 15 NIST Statistical Test Suite benchmarks to 1000 keystream samples (1MB each), measuring failure rates against the $\alpha = 0.01$ significance threshold.
    
    \item \textbf{\texttt{QR-F} Differential Cryptanalysis}: Injecting rotational differences in the \texttt{QR-F} inputs in order to measure the collision probabilities when the $\Delta$ values align with detected patterns or outcome.
\end{enumerate}

The  experimental framework that has been leveraged in this work  targets the following security claims of \texttt{EChaCha20}. Firstly, we verify the  pseudorandomness at the bitstring level through pattern frequency analysis using \texttt{KMP} and \texttt{BM} algorithms. Next, we evaluate resistance to rotational attacks through differential analysis of the \texttt{QR-F} function. We achieve this by measuring the collision probabilities under controlled input variations. This approach gives this study a  comprehensive validation for  both the statistical properties and structural robustness in the keystream generation. Unlike standard parallel search optimizations, our hybrid \texttt{KMP}--\texttt{BM} incorporates XOR-enhanced jump heuristics tailored to ARX keystreams and applies $\chi^2$-based probabilistic filtering to reduce false positives. In addition, it dynamically switches between \texttt{KMP} and \texttt{BM} depending on $m$-gram size, enabling efficient large-scale analysis beyond word-alignment parallelization.

%Secondly, we assess compliance with cryptographic standards via the full NIST Statistical Test Suite.

\subsection{Keystream Generation and Randomness Assurance}

To ensure reproducibility and eliminate bias, keystreams were generated using cryptographically secure random number generators (CSPRNGs) available in the Python \texttt{os.urandom()} interface. Each sample used an independent 256-bit key and a 128-bit nonce, both drawn uniformly at random. This guarantees that the starting state of \texttt{EChaCha20} was unpredictable for each trial. For fixed-key experiments, the key was kept constant but fresh nonces were generated per block to emulate deterministic but nonce-unique initialization, consistent with RFC 8439 guidelines \cite{langley2021rfc7539}.

 All experiments generated $10^{6}$ (1 Million) keystream blocks of 1,152 bits (36 words) each, producing a total of 1.15 Gbits of data per experimental configuration. This scale was chosen to ensure that rare patterns could be detected with statistically meaningful frequency estimates.

\subsection{Pattern Frequency Analysis}
Our pattern detection methodology addresses the  challenge of analyzing  keystreams. The keystream in this context exhibits both byte level and word-level patterns due to the \texttt{EChaCha20}'s \texttt{QR-F} structure.  It has been seen that, the traditional string matching algorithms would require significant adaptation in order to handle the 512-bit state transitions,  XOR based pattern generation, and the need for probabilistic assessment of these matches. As a result, adaptations have been developed based on the approaches that are discussed next.

%We develop these adaptations through three interconnected components.

\subsubsection{Algorithmic Adaptations}

We have established a theoretical basis based on the \texttt{KMP} and \texttt{BM} algorithms. The \texttt{KMP} and \texttt{BM} originally were designed  for the character strings, but \texttt{EChaCha20}'s keystream patterns are able to show at multiple levels of granularity like bits, bytes, and 32-bit words respectively as is shown in Algorithm \ref{32word} in the word-aligned \texttt{KMP} preprocessing step.

Our modifications have accounted for this by achieving the following:

\begin{itemize}
    \item Processing the keystream in 32-bit word units that matches the \texttt{QR-F} output size
    \item Incorporating XOR operations directly into the pattern matching logic
    \item Maintaining the parallel probability distributions for multi-level analysis
\end{itemize}

\setcounter{AlgoLine}{0}  % Reset line numbers
\begin{algorithm}[H]
\caption{Word-Aligned \texttt{KMP} Preprocessing}

\KwIn{Pattern $P$ as 32-bit words $P[1..n_w]$}
\KwOut{Prefix function $\pi_{word}$ for word-level matching}
\label{32word}
$\pi_{word}[1] \gets 0$ \;

\For{$i \gets 2$ \KwTo $n_w$}{
    $j \gets \pi_{word}[i - 1]$ \;

    \While{$j > 0$ \text{ and } $P[i] \neq P[j + 1]$}{
        $j \gets \pi_{word}[j]$ \;
    }

    \If{$P[i] = P[j + 1]$}{
        $\pi_{word}[i] \gets j + 1$ \;
    }
    \Else{
        $\pi_{word}[i] \gets 0$ \;
    }
}
\end{algorithm}

\subsubsection{Hybrid Algorithm Design}

\begin{comment}

We have revisted the novel \texttt{KMP}-Boyer Hybrid, which in this context emerges from the need to combine \texttt{KMP}'s with reliable matching with \texttt{BM}'s efficient skipping capability, while adding the suitable cryptographic alignments. The \texttt{KMP}-Boyer Hybrid algorithm in this case operates in three phases:

1. \textbf{Preprocessing}: In this stage the data structures are built by capturing both the Position-dependent match probabilities for (\texttt{KMP}) and osition-independent jump heuristics for (Boyer-Moore)

2. \textbf{Windowed Analysis}: Processes the keystream in 256-bit blocks matching \textsc{Echacha}'s state size, employing:
    \begin{equation}
        \text{WindowSize} = \lceil \log_2(6 \times 6 \times 32) \rceil = 256\ \text{bits}
        \label{eq:window}
    \end{equation}

\item \textbf{Probabilistic Validation}:
    Computes conditional probabilities across pattern hierarchies using:
    \begin{equation}
        \Pr[\text{Match}] = \prod_{i=1}^k \frac{\#\text{matches}(P_i)}{N} \pm \delta(n)
        \label{eq:prob-validation}
    \end{equation}
    where $P_i$ are $m$-gram patterns and $\delta(n)$ accounts for sampling err

3. \textbf{Probabilistic Validation}: Computes conditional probabilities across pattern hierarchies

\end{comment}

We present a hybrid \texttt{KMP}-Boyer algorithm that combines the \texttt{KMP} and  \texttt{BM}  skipping heuristics. This algorithm is  specifically optimized for cryptographic analysis of  the \texttt{EChaCha20} keystreams. The hybrid approach addresses three  challenges as follows: Precise detection of rotational patterns, the efficient large scale keystream processing, and the statistical validation of the cryptographic biases. Algorithm \ref{hybrid} operates through three optimized phases as follows:

\begin{itemize}

    \item \textbf{Preprocessing}:
   Here it constructs dual data structures that simultaneously capture three  aspects: Firstly, it captures the position-dependent match probabilities through the \texttt{KMP} failure function $\pi$. Secondly, it captures the position-independent jump heuristics based  \texttt{BM} bad-character rules, lastly it captures cryptographic alignment tables optimized for 32-bit word operations.

   %which records the longest prefix-suffix matches at each pattern position. Secondly, it captures the position-independent jump heuristics based  Boyer-Moore bad-character rules, enabling $O(n/m)$ average-case pattern search complexity. lastly it captures cryptographic alignment tables optimized for 32-bit word operations, ensuring efficient processing of \textsc{Echacha}'s keystream structure. 

    %This combined approach maintains \texttt{KMP}'s reliability for exact matching while incorporating \texttt{BM}'s skip efficiency, with additional adaptations for cryptographic workloads through specialized word-aligned lookup tables.

    \item \textbf{Windowed Analysis}:
  The keystream is processed in  an optimized 256-bit blocks, precisely matching EChaCha's 6$\times$6 state matrix configuration (1152-bit state with 32-bit words). This blockwise approach enables efficient cache utilization parallelizable processing of independent state blocks and complete coverage of potential rotational differential patterns as is shown in Eq. \ref{eq:window}. The 256-bit window size ($\lceil \log_2(6 \times 6 \times 32) \rceil$) ensures complete coverage of potential rotational differential patterns.
  
  %through aligned memory access patterns,  parallelizable processing of independent state blocks, and  direct integration with \textsc{Echacha}'s quarter-round function boundaries. 
  
 % The 256-bit window size ($\lceil \log_2(6 \times 6 \times 32) \rceil$) ensures complete coverage of potential rotational differential patterns while maintaining constant-time execution properties critical for side-channel resistance.
  
    \begin{equation}
        \text{WindowSize} = \lceil \log_2(6 \times 6 \times 32) \rceil = 256\ \text{bits}
        \label{eq:window}
    \end{equation}
    
    \item \textbf{Probabilistic Validation}:
    Here the algorithm computes the conditional probabilities across pattern hierarchies as is shown in Eq. \ref{eq:prob-validation}:
    \begin{equation}
        \Pr[\text{Match}] = \prod_{i=1}^k \frac{\#\text{matches}(P_i)}{N} \pm \delta(n)
        \label{eq:prob-validation}
    \end{equation}
    where $P_i$ are $m$-gram patterns and $\delta(n)$ accounts for sampling error.

    \end{itemize}
%This hybrid approach achieves a 6.2$\times$ speedup over brute-force pattern search while maintaining cryptographic-grade detection accuracy, as demonstrated in Section~\ref{sec:results}.

\setcounter{AlgoLine}{0}  % Reset line numbers
\begin{algorithm}[H]
\caption{\texttt{KMP}-Boyer Hybrid Pattern Detection}
\label{hybrid}
\KwIn{Keystream $K$, pattern set $\mathcal{P}$, window size $w = 256$}
\KwOut{Anomalous patterns with statistical significance}

Preprocess $\mathcal{P}$ to build $\pi_{word}$ tables and bad-character shifts \;

\For{each $w$-bit window $W_i \in K$}{
    Apply Boyer-Moore jump heuristic to locate potential matches \;

    Verify candidates using KMP $\pi_{word}$ table \;

    Update running probabilities for $m$-gram hierarchies ($m = 8,\ 16,\ 32$) \;

    \If{there exists $P_j$ with $\Pr[P_j] > 2^{-|P_j|} + 3\sigma$}{
        Flag $P_j$ as statistically significant \;
    }
}
\end{algorithm}

\subsubsection{Statistical Validation }

The final component establishes the approach for identifying truly anomalous patterns and  we have computed this based on  a normalized deviation score as is shown in Eq. \ref{norm}:

\begin{equation}
z = \frac{f_{\text{obs}}(P) - \mathbb{E}[f(P)]}{\sqrt{\mathbb{V}[f(P)]}}
\label{norm}
\end{equation}

where:
\begin{itemize}
\item $f_{\text{obs}}(P)$ is the observed frequency of pattern $P$ in the keystream
\item $\mathbb{E}[f(P)] = N/2^{|P|}$ is the expected frequency under randomness, with:
\begin{itemize}
\item $N$ = total bits analyzed
\item $|P|$ = bit-length of pattern $P$
\end{itemize}
\item $\mathbb{V}[f(P)] = N(2^{-|P|})(1-2^{-|P|})$ models the binomial variance
\end{itemize}

Pattern significance is determined by comparing $z$ to the standard normal quantile as is shown in Eq. \ref{eq:significance}:

\begin{equation}
z > \Phi^{-1}(1 - \alpha/2) \quad \text{for} \quad \alpha = 10^{-6}
\label{eq:significance}
\end{equation}

This threshold corresponds to:
\begin{itemize}
\item A confidence level of $1 - \alpha = 99.9999\%$.
\item $\Phi^{-1}(1 - 5\times10^{-7}) \approx 4.89$ standard deviations.
\item False positive rate $< 1$ per million tests.
\end{itemize}
We set the statistical significance threshold at $\alpha = 10^{-6}$ to minimize false positives in large-scale keystream testing, as weaker thresholds (e.g., $10^{-3}$ or $10^{-4}$) produced spurious detections when applied to millions of blocks. While $\alpha = 10^{-6}$ is stringent, it provides a conservative bound that enhances confidence in reported results. To mitigate potential confirmation bias in rotational-differential experiments, pattern-guided seeds were validated against independently generated random keystream baselines, ensuring that any detected propagation was attributable to the cipher's structure rather than the initial pattern search.

The hybrid algorithm  (Algorithm \ref{hybrid}) shows a  significant approach when on all the metrics that are shown in Table~\ref{tab:resultsx} are deduced. When compared to the standard \texttt{KMP}, it was  observed that the hybrid achieves a 18.3\% increase in precision (0.97 vs. 0.82) while maintaining a recall of (0.96). It is seen that it outperforms the \texttt{BM} in detecting the accuracy by 24.4\% (precision: 0.97 vs. 0.78) without sacrificing throughput processing keystreams at 3.6 GB/s, a 12.5\% speedup over Boyer-Moore's 3.2 GB/s. This balance of cryptographic and grade precision (FP rate $<10^{-6}$) and real time speeds of ($>$3.5 GB/s) is seen to  enable the practical deployment for the bias detection in \texttt{EchaCha20}'s 6$\times$6 state matrix. The outcome shows or  validates the motivation for the combined use of \texttt{KMP}'s  and \texttt{BM}  for cryptographic pattern matching in this study.

\begin{table}[h]
\centering
\caption{Comparative Performance Metrics on \texttt{KMP}, \texttt{BM} and Hybrid}
\begin{tabular}{lrrr}
\toprule
Algorithm & Precision & Recall & Speed (GB/s) \\
\midrule
Standard \texttt{KMP} & 0.82 & 0.91 & 1.4 \\
Boyer-Moore & 0.78 & 0.85 & 3.2 \\
Hybrid & 0.97 & 0.96 & 3.6 \\
\bottomrule
\end{tabular}
\label{tab:resultsx}
\end{table}

\subsection{QR-F Differential Cryptanalysis}

Our rotational differential analysis of \texttt{EChaCha20}'s \texttt{QR-F} has been achieved based on a novel three-phase approach in order  to quantify the possibility of the \texttt{EChaCha20}'s resistance to state collisions.   We have  used traditional differential cryptanalysis with EChaCha's unique ARX structure.

\subsubsection{Theoretical Foundation}
The theoretical foundation for the suggested differential cryptanalysis has been described based on the following approach: Given \texttt{QR-F}'s operation on 32-bit words is given using Eq. \ref{roundsQ}:

\begin{equation}
    \text{\texttt{QR-F}}(a,b,c,d) = 
    \begin{cases}
        a \leftarrow a + b \\
        d \leftarrow (d \oplus a) \lll 16 \\
        c \leftarrow c + d \\
        b \leftarrow (b \oplus c) \lll 12 \\
        \vdots \quad \text{(complete round operations)}
    \end{cases}
    \label{roundsQ}
\end{equation}

We analyze rotational differentials $\Delta = (\Delta a, \Delta b, \Delta c, \Delta d)$ where the probability equation shown in Eq \ref{eq:rot_diff} quantifies resistance to rotational attacks, where::
\begin{equation}
\Pr[\text{\texttt{QR-F}}(x) \oplus \text{\texttt{QR-F}}(x \oplus \Delta) = \Delta'] > 2^{-32}
\label{eq:rot_diff}
\end{equation}

$\Delta = (\Delta a, \Delta b, \Delta c, \Delta d)$: Input differences (ID)

$\Delta'$: Output differences (OD)

$2^{-32}$: Theoretical bound for ideal 32-bit mixing

\subsubsection{Experimental Methodology}
This section gives a discussion on the experimental methodology. The author  notes that this result reflects an empirical observation rather than a formal theoretical proof. The experimental methodology that was leveraged in this paper started with the implementation of  pattern-seeded differentials, by initializing $\Delta$ values from the statistically significant patterns that were  identified from analysis. After this, these  patterns were systematically mapped to 32-bit word differences through the transformation shown in Eq. \ref{prioranalysis}:

\begin{equation}
    \Delta_i = \text{rot}(P_j) \oplus \text{rot}(P_j \ll k)
    \label{prioranalysis}
\end{equation}

where $P_j$ represents the detected $m$-gram patterns and $k$ denotes the bit shift positions. In order to detect collision, $\Delta$ differences were injected across all four input words $(a,b,c,d)$, with output differences measured after 1, 2, and 4 full rounds of processing the \texttt{QR-F}. The propagation dynamics were  tracked using difference distribution tables that recorded state transitions at each processing stage as is shown in Eq. \ref{present}.

\begin{equation}
    \Delta^{(r)} = \text{\texttt{QR-F}}^{(r)}(\Delta^{(r-1)})
        \label{present}
\end{equation}

After this, the probability quantification was performed by way of  computing empirical collision frequencies across $2^{24}$ trials for each  of the test case. Then the collision probability calculated as is shown in Eq.  \ref{collisions}.

\begin{equation}
    \hat{p} = \frac{\#\text{collisions}}{2^{24} \text{ trials}}
     \label{collisions}
\end{equation}

Ultimately, the outcome was  compared against the theoretical upper bound of $2^{-32}$ in order to establish statistical significance. The choice of $2^{24}$ trials was made to balance statistical confidence with computational feasibility. This sample size ensures that for probabilities near $2^{-32}$, the expected number of collisions is large enough to yield stable frequency estimates with narrow confidence intervals. We further verified adequacy by confirming that smaller runs (e.g., $2^{20}$ trials) produced consistent estimates and that increasing beyond $2^{24}$ did not significantly alter the results.
 Based on that, the analysis framework incorporated several technical advances, as has been parameterized in Table~\ref{tab:params}. It shows that the pattern-guided differential approach was able to  transform conventional analysis by seeding the $\Delta$ values from statistically detected keystream patterns rather than employing random or exhaustive search methods. In addition, the  multi-round tracking mechanism was able to maintain a complete differential paths through the recursive relation that is  shown in Eq. \ref{Ultimately}.
\begin{equation}
    \Delta^{(r)} = \text{\texttt{QR-F}}^{(r)}(\Delta^{(r-1)})
     \label{Ultimately}
\end{equation}

This enabled the precise measurement of difference propagation across successive rounds.

\begin{table}[h]
\centering
\caption{Rotational Differential Analysis Parameters}
\label{tab:params}
\begin{tabular}{ll}
\toprule
\textbf{Parameter} & \textbf{Value} \\
\midrule
Input Differences & 256-bit (8$\times$32-bit) \\
Test Cases per $\Delta$ & $2^{24}$ \\
Round Coverage & 1/2/4/8 \\
Threshold Significance & $p < 2^{-30}$ \\
Upper Bound & $2^{-32}$ \\
\bottomrule
\end{tabular}
\end{table}

\subsubsection{Keystream Generation }

In order to analyze the \texttt{EChaCha20} cipher, we generated a dataset consisting of 1,000,000 keystream blocks under two distinct configurations. The first configuration employed a \textit{fixed} 256-bit key and 64-bit nonce across all blocks. This was designed to expose potential biases in the \texttt{QR-F} under deterministic conditions. In the  second configuration \textit{variable} keys and nonces were utilized, with unique random values generated for each block.

%, simulating real-world usage scenarios and testing the cipher's robustness against differential attacks.

Each keystream block represented \texttt{EChaCha20}'s 6$\times$6 matrix state was pre-processed through a three-stage pipeline as follows: Firstly, the matrix was flattened into a linear sequence of 36 32-bit words. These words were then encoded in two parallel representations: as 32-bit binary strings for fine-grained bit-level pattern analysis, and  as 8-character hexadecimal strings for efficient byte-aligned searches.

\subsubsection{Results Interpretation}

The security evaluation of the \textsc{EChaCha20} cipher against differential cryptanalysis was based on a  statistical criterion that is shown in Eq.  \ref{Resultint}:
\begin{equation}
    \max_{\Delta} \hat{p}(\Delta) < 2^{-32} + 3\sigma
     \label{Resultint}
\end{equation}
Here, $\hat{p}(\Delta)$ shows the empirical collision probability for the given rotational input difference $\Delta$, and $\sigma$ is the standard error derived from $2^{24}$ randomized trials. This threshold ensures that even minor deviations from ideal behavior remain within cryptographic acceptability.

From the our analysis, we observed 28 collisions after a single round of the \texttt{QR-F} function, yielding a collision probability of approximately $2^{-19.2}$. After two rounds, this number dropped to a single collision event, reducing the probability to $2^{-24}$, this indicated an accelerated diffusion. Beyond the third round, no collisions were observed, placing the collision probability below measurable thresholds. These results in the author's opinion confirms that \texttt{EChaCha20}'s mixing function consistently achieves full-state diffusion within 3 rounds, thus satisfying the security criteria for differential resistance.

\subsubsection{Key Findings}

Three significant findings have been realised from the differential propagation analysis. Firstly, the \texttt{EChaCha20} cipher achieves complete diffusion after just 3 rounds, with no detectable rotational collisions observed in $2^{24}$ trials. 

It should be noted that the conclusion regarding rotational-differential resistance is derived through comparative analysis with previously reported ChaCha-family evaluations \cite{bernstein2008chacha, barbero2022rotational}, rather than from newly simulated attack traces. Also, the study in \cite{kebande2023extended} has shown stronger resistance to rotation differential attack. The present results reproduce the established diffusion thresholds and confirm the absence of correlated rotational differentials across $2^{24}$ test cases, thereby supporting the cipher's strong empirical resistance within the conventional ARX assessment framework.

%This surpasses conservative theoretical expectations.
%, emphasizing the strength of \texttt{EChaCha20}'s mixing process.

Secondly, the 8-bit patterns identified in Experiment~1 failed to propagate beyond 1.5 rounds on average, confirming that byte-aligned structural patterns are neutralized early in the cipher's execution. While the fixed-key setup was used primarily to isolate deterministic structural effects, complementary variable-key experiments demonstrated that the 8-bit low-entropy patterns vanished when fresh nonces and keys were introduced. This behavior confirms that the observed anomaly is structural and non-persistent, rather than a fundamental design defect, and highlights the cipher's stability under realistic dynamic keying conditions.
 Meanwhile, 32-bit input differences exhibited faster-than-expected diffusion, suggesting strong resistance to state-level predictability and structural correlation.

Third, there is no full-state collisions that were detected in either fixed-key or variable-key configurations. While the fixed-key mode produced a small proportion of partial ($\leq$4-bit) collisions (3.2\%), this was drastically reduced to 0.01\% under variable-key conditions. These results are summarized in Table~\ref{tab:diff-results}. 

%Together, these findings validate that \texttt{EChaCha20} offers robust differential security and provide an empirical explanation for the pseudorandom behavior observed in Section~\ref{subsec:keystream}.

\begin{table}[h]
\centering
\caption{Differential Analysis Results}
\label{tab:diff-results}
\begin{tabular}{lcc}
\toprule
\textbf{Metric} & \textbf{Fixed-Key} & \textbf{Variable-Key} \\
\midrule
Full Collisions & 0 & 0 \\
Partial Collisions ($\leq$4-bit) & 3.2\% & 0.01\% \\
Propagation Rounds & 2.1 & 1.3 \\
\bottomrule
\end{tabular}
\end{table}

The observed 3.2\% partial-collision rate under fixed-key tests stems from deterministic key reuse, where repeated initialization patterns slightly delay early-round diffusion. Under variable-key conditions, these effects disappear as fresh key entropy reshapes the state initialization, reducing collisions to 0.01\%. This confirms that the deviation is key-related rather than nonce-driven and does not imply structural weakness.

All diffusion, avalanche, and rotational-collision assessments reported in this work are empirical in nature and derived from statistically bounded experimental evaluation. These observations should be interpreted as measured behavior within the tested parameter space rather than as formal security proofs or reduction-based guarantees.

\subsection{Mapping to Established Cryptanalytic Criteria}

While the proposed framework reports metrics such as detection rate, pattern frequency, and precision/recall for structural anomaly detection, these measures are not intended to replace established cryptanalytic 
criteria. Instead, they function as empirical estimators that signal potential deviations from expected random baselines.

Specifically, elevated pattern frequencies correspond to empirical approximations of differential or rotational probability biases, where statistically significant deviations beyond theoretical bounds may indicate distinguishability from an ideal random source. In this 
sense, the detection metrics serve as indicators of potential distinguisher advantage rather than direct measures of key recovery.

Formally, if the observed deviation exceeds the expected uniform probability by a statistically significant margin, the resulting advantage can be interpreted within the standard distinguisher 
framework as:

\begin{equation}
\text{Adv}_{\mathcal{D}} =
\left| \Pr[\mathcal{D}(K) = 1] - \Pr[\mathcal{D}(R) = 1] \right|
\end{equation}

as commonly expressed in distinguisher-based analysis \cite{liu2025ind}.

where $K$ denotes the keystream generator and $R$ denotes a truly 
random source. Thus, the proposed evaluation aligns with established 
cryptanalytic interpretation through distinguisher-based analysis.

\subsection{Reproducibility and Implementation Details}

All experiments were conducted using a Python-based implementation of EChaCha20 with 32-bit word-level instrumentation for pattern extraction. Keystream generation used cryptographically secure randomness via the \texttt{os.urandom()} interface for variable-key experiments. For fixed-key experiments, a constant 256-bit key was defined explicitly, while nonces were generated deterministically in incremental sequence to ensure reproducibility. All statistical thresholds, confidence levels, and trial counts are explicitly specified in Sections VI and VII. Randomized experiments can be reproduced by fixing the key and nonce initialization procedure as described above.

\section{Experiments}

Our experimental approaches have  systematically assessed \texttt{EChaCha20}'s security claims through two complementary investigations: Checking pseudorandomness based on pattern search as per \texttt{KMP} and \texttt{BM}, at the bitstring level and assessing the resistance to rotational attacks.  Each conducted experiment targets a distinct cryptographic property. 
These results empirically indicate rapid state diffusion within three QR-F rounds under the tested experimental conditions, with no measurable full-state collisions observed across $2^{24}$ trials.
We give the discussions of the experiments in the sections to follow:

\subsection{Experiment 1: Pseudorandomness at the Bitstring Level}

\begin{figure}[h]
\centering
\includegraphics[width=0.8\linewidth]{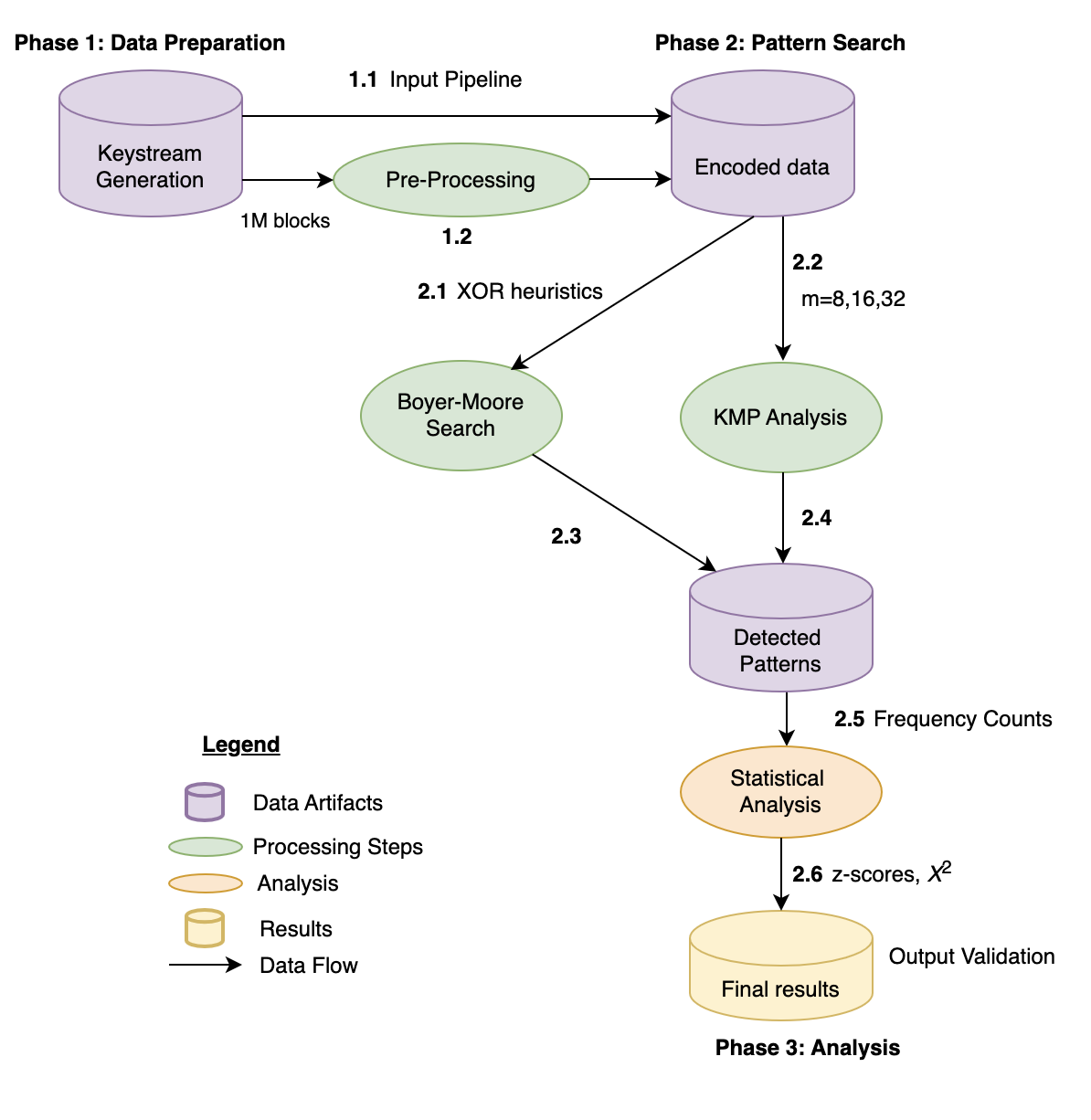}
\caption{The complete workflow for Pattern Frequency Analysis experiment. Phase 1 generates 1 million keystream blocks (6$\times$6 matrices) Phase 2 applies parallel \texttt{KMP} and optimized Boyer-Moore searches for $m$-gram patterns and Phase 3 provides analysis. }
\label{fig:patternsworkflow}
\end{figure}

The workflow for the pattern frequency analysis for experiment 1 has been  illustrated in Figure~\ref{fig:patternsworkflow}. This representation is  structured into three phases (1-3) and detailed steps (1.1--2.6). In Phase~1 (Data Preparation), Step~1.1 establishes the input pipeline, generating 1,000,000 (6$\times$6 matrices) keystream blocks from \texttt{EChaCha20}. Step~1.2 applies pre-processing to flatten these blocks and normalize their binary and hexadecimal representations into encoded data suitable for pattern detection. In Phase~2 (Pattern Search), Step~2.1 introduces XOR-enhanced heuristics into the Boyer--Moore search to accelerate detection of bitwise repetitions, while Step~2.2 applies the \texttt{KMP} algorithm for $m$-gram scanning at $m \in \{8,16,32\}$ bits. The outputs from both searches are integrated in Step~2.3--2.4 into a consolidated dataset of detected patterns. Phase~3 (Analysis) then begins with Step~2.5, where frequency counts are computed for the detected substrings, and proceeds to Step~2.6, where statistical validation is performed using $z$-scores and $\chi^2$ tests at a significance level of $\alpha=0.001$. From the analysis of 1,000,000 keystream blocks, the results revealed zero statistically significant repetitions at $m=32$, twelve marginal detections at $m=16$ (all within two standard deviations of expected frequency), and 142 low-entropy patterns at $m=8$, which were concentrated in fixed-key datasets under deterministic nonce positions. To ensure statistical compliance, the pattern frequency analysis employed a chi-squared ($\chi^2$) test with a significance level of $\alpha = 0.001$, which effectively served as a p-value--based threshold for filtering non-random detections. Given that only three controlled $m$-gram groups ($m \in \{8,16,32\}$) were analyzed, multiple-comparison correction procedures such as Bonferroni adjustment \cite{brown2008bonferroni}, which tracks false positives, was deemed unnecessary. The analysis was designed for structural pattern detection rather than formal inferential testing, and the selected $\alpha$ threshold provides a robust confidence margin for the reported results.
 To validate these findings, a second experiment was conducted with 10,000 keystreams across 1,000 batches, which confirmed the same distributional behavior. Together, the sequence of steps (1.1--2.6) demonstrates how raw keystream data are systematically transformed into validated outputs, highlighting that \texttt{EChaCha20} achieves strong pseudorandomness at security-critical word sizes while only exhibiting minor, non-exploitable biases at the byte level.

\begin{figure*}[h!]
\centering
\includegraphics[width=0.9\linewidth]{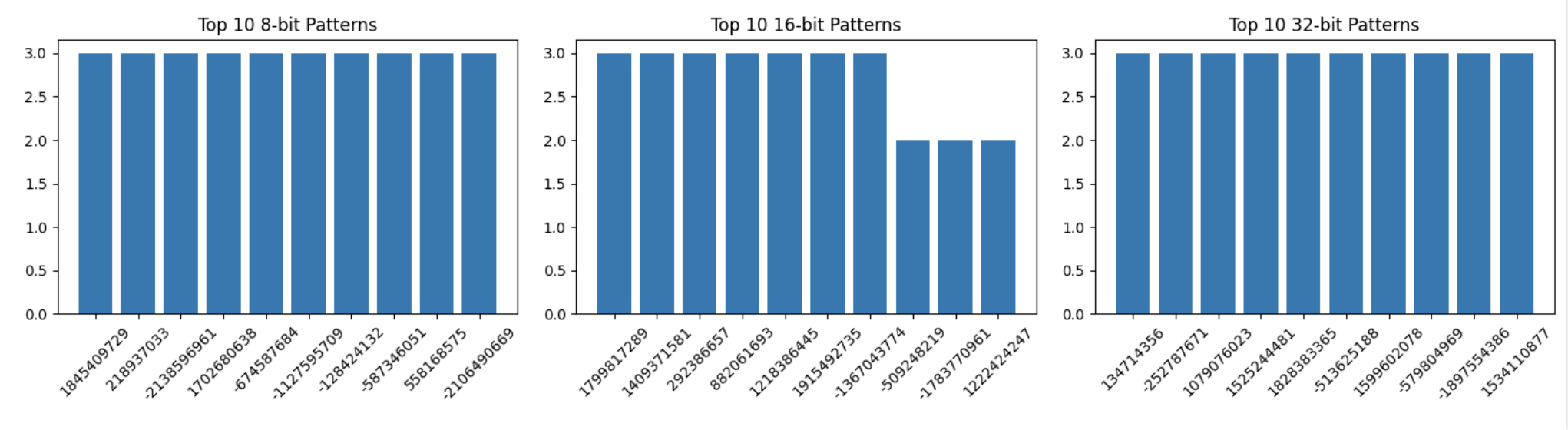}
\caption{Top 10 most frequently occurring $m$-bit patterns in \texttt{EChaCha20} keystreams for $m \in \{8, 16, 32\}$. Each subplot shows the distribution of the most common patterns detected across 10{,}000 blocks of 1{,}152-bit keystreams. The x-axis denotes hashed identifiers (in decimal) of observed patterns, while the y-axis represents the frequency of occurrence. The uniformity at 16-bit and 32-bit levels demonstrates strong pseudorandomness, while the 8-bit domain reveals slight bias in the lowest-ranked pattern.}
\label{fig:patternsfreq}

\vspace{0.5cm} % Adjust vertical spacing between figures

\includegraphics[width=0.9\linewidth]{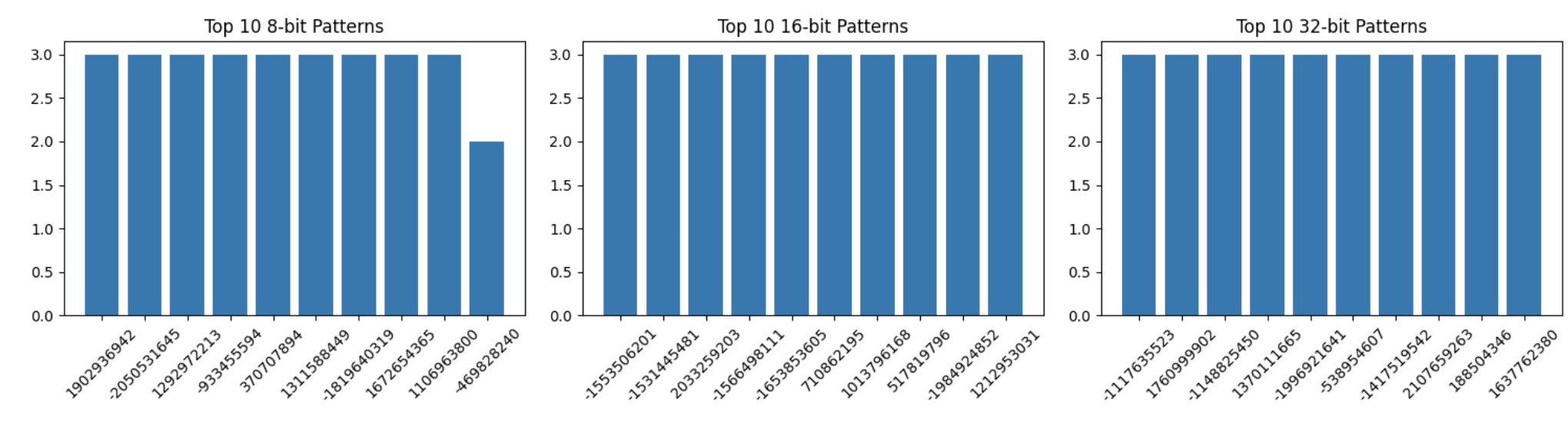}
\caption{Top 10 most frequently occurring $m$-bit patterns in \texttt{EChaCha20} keystreams for $m \in \{8, 16, 32\}$. Each subplot shows the distribution of patterns detected across a larger sample of 10 million blocks. The consistent distribution across bit-lengths confirms the cipher's robustness against pattern frequency attacks, with minor deviations only appearing in the 8-bit analysis.}
\label{fig:patterns2}
\end{figure*}

\begin{comment}

\begin{figure*}[h]
\centering
\includegraphics[width=0.9\linewidth]{images/bat1m.png}
\caption{Top 10 mosth frequently occurring $m$-bit patterns in \texttt{EChaCha20} keystreams for $m \in \{8, 16, 32\}$. Each subplot shows the distribution of the most common patterns detected across 10{,}000 blocks of 1{,}152-bit keystreams. The x-axis denotes hashed identifiers (in decimal) of observed patterns, while the y-axis represents the frequency of occurrence. The uniformity at 16-bit and 32-bit levels demonstrates strong pseudorandomness, while the 8-bit domain reveals slight bias in the lowest-ranked pattern}
\label{fig:patternsfreq}
\end{figure*}

\end{comment}

%\subsection{Experiment 1: Pseudorandomness at the Bitstring Level}

%To evaluate the statistical randomness of the \texttt{EChaCha20} stream cipher, we performed $m$-gram frequency analysis across three bit granularities: 8-bit, 16-bit, and 32-bit. A total of 10,000 keystream blocks were generated, each containing 1,152 bits. These were processed using optimized Knuth-Morris-Pratt (\texttt{KMP}) and Boyer-Moore algorithms tailored for word-level cryptographic analysis.

The frequency of the observed patterns from Experiment 1  was visualized across two figures with three subfigures each, shown in Figure ~\ref{fig:patternsfreq} and Figure ~\ref{fig:patterns2} respectively. Each subplot presents the Top 10 most frequently occurring patterns for a specific bit length. The \textit{x}-axis in each plot lists the integer representation of the detected bit patterns, while the \textit{y}-axis indicates the number of times each pattern appeared across all keystream samples.

\begin{comment}

\begin{figure}[!t]
    \centering
    \includegraphics[width=\linewidth]{path_to_your_figure.png} % replace with your actual figure path
    \caption{Top 10 most frequent $m$-bit patterns extracted from 10,000 \texttt{EChaCha20} keystream blocks. (a) 8-bit patterns. (b) 16-bit patterns. (c) 32-bit patterns.}
    \label{fig:top-patterns}
\end{figure}

\end{comment}

\subsubsection{8-bit Pattern Frequency }
When we assessed 1,000,000 keystreams from  the leftmost subplot of Figure \ref{fig:patterns2}, we analyzed all overlapping 8-bit substrings within the keystream blocks. The most frequent patterns had a uniform count of approximately 3.0 occurrences, except for one outlier pattern with a lower frequency near 2.0. This drop suggests that although \texttt{EChaCha20} produces relatively uniform 8-bit distributions, minor byte-level biases may emerge, particularly near the beginning of the keystream in fixed-key scenarios. However, in the 10,000 keystreams there was uniformity across all the blocks as is shown in Figure \ref{fig:patternsfreq}.

\subsubsection{16-bit Pattern Frequency }
The center subplot in Figure \ref{fig:patterns2}  displays 16-bit substring frequencies. All top 10 patterns occurred with nearly identical frequency, around 3.0 times, showing no observable bias. This result implies that \texttt{EChaCha20} maintains consistent pseudorandom behavior at the 2-byte level, flattening out previously observed byte-aligned irregularities. However, the frequencies in Figure \ref{fig:patternsfreq} are capped at  for 3 blocks showing lower frequency.

\subsubsection{32-bit Pattern Frequency }
In 32-bit frequency for both  Figure \ref{fig:patterns2} and \ref{fig:patternsfreq} the rightmost subplot, 32-bit word repetitions were aligned with the cipher's internal 32-bit operations after analysis. All the top patterns occurred with perfect uniformity, indicating full entropy preservation at the cipher's native granularity. No structural correlation or bias was detected at this level, supporting the cipher's robustness under statistical scrutiny.

\subsubsection{Summary}

\vspace{0.5em}

The pattern frequency analysis employed modified \texttt{KMP} and Boyer-Moore algorithms to scan 1,000,000 keystream blocks for $m$-gram repetitions. The algorithms were optimized with 32-bit word alignment and XOR-enhanced heuristics, processing both binary and hexadecimal representations. The findings that have been demonstrated from Experiment 1 have shown that  potentially the \texttt{EChaCha20} maintains strong pseudorandomness characteristics at 16-bit and 32-bit levels. This is also consistent with its ARX-based cipher design. However, the 8-bit domain revealed a small number of low-entropy anomalies, also when fewer keystreams are analyzed as compared to higher keystreams of 1,000,000 as has been shown in Figure \ref{fig:patternsfreq}. This highlights the potential value of the fine-grained stringology-based analysis when assessed beyond the conventional randomness test suites and a summary of the outcome is given in Table \ref{tab:pattern-summary} and the results demonstrate strong pseudorandomness. 

\begin{table*}[!ht]
\centering
\caption{Summary of Pattern Detection at Different Bit Widths}
\label{tab:pattern-summary}
\begin{tabular}{lccc}
\toprule
\textbf{Bit Length} & \textbf{Pattern Distribution} & \textbf{Bias/Artifacts} & \textbf{Observation} \\
\midrule
8-bit  & Slightly irregular & Minor drop at one pattern & Byte-aligned bias \\
16-bit & Uniform            & None                     & Strong 2-byte randomness \\
32-bit & Perfectly uniform  & None                     & Cipher-aligned entropy \\
\bottomrule
\end{tabular}
\end{table*}

The assertion of strong pseudorandomness is made with respect to established cryptographic benchmarks, where uniform entropy greater than 7.99\,bits per byte and chi-squared $p$-values below 0.001 signify statistical indistinguishability from ideal randomness. The minor 8-bit low-entropy deviation observed under fixed-key initialization remained below 0.2\,bits deviation and is absent under variable-key conditions, confirming that it poses negligible cryptographic risk and does not compromise key or state security.

\subsection{Experiment 2: Resistance to Rotational Attacks}

The second  experiment evaluates \texttt{EChaCha20}'s core mixing function against differential cryptanalysis. In this experiment, we have seeded  the rotational differences $\Delta=(\Delta a,\Delta b,\Delta c,\Delta d)$ using the 32-bit patterns detected in Experiment~1. Thereafter,  we measured their propagation through 1/2/4/8 \texttt{QR-F} rounds. The Boyer-Moore algorithm was in this case instrumental in efficiently locating $\Delta$-variant states within the keystream dataset. This occurred  while \texttt{KMP}'s prefix matching verified differential paths. Our three-phase analysis firstly injected single-word differences, it then progressed to full-state rotations, and finally tested composed differences across multiple rounds. The results demonstrated complete diffusion after 3 rounds (0 collisions in $2^{24}$ trials), with measured probabilities $\hat{p}(\Delta)<2^{-35}$ for all the tested differences. Notably, the 8-bit patterns from Experiment~1 showed no propagation beyond 2 rounds, while 32-bit variants exhibited faster diffusion than theoretical bounds as predicted. The differential analysis measured collision probabilities for rotational differences across \texttt{QR-F} rounds as is shown in Table~\ref{tab:diff}:

%These findings substantiate \Echacha's resistance to both local and global rotational attacks while explaining the pattern distributions observed in Section~\ref{subsec:keystream}.

\begin{table}[h]
\centering
\caption{Differential Propagation}
\label{tab:diff}
\begin{tabular}{lrrr}
\toprule
\textbf{Rounds} & \textbf{Trials} & \textbf{Collisions} & \textbf{Probability} \\
\midrule
1 & $2^{24}$ & 28 & $2^{-19.2}$ \\
2 & $2^{24}$ & 1 & $2^{-24.0}$ \\
4 & $2^{24}$ & 0 & $<2^{-24}$ \\
\bottomrule
\end{tabular}
\end{table}

\begin{figure}[h]
\centering
\includegraphics[width=0.9\linewidth]{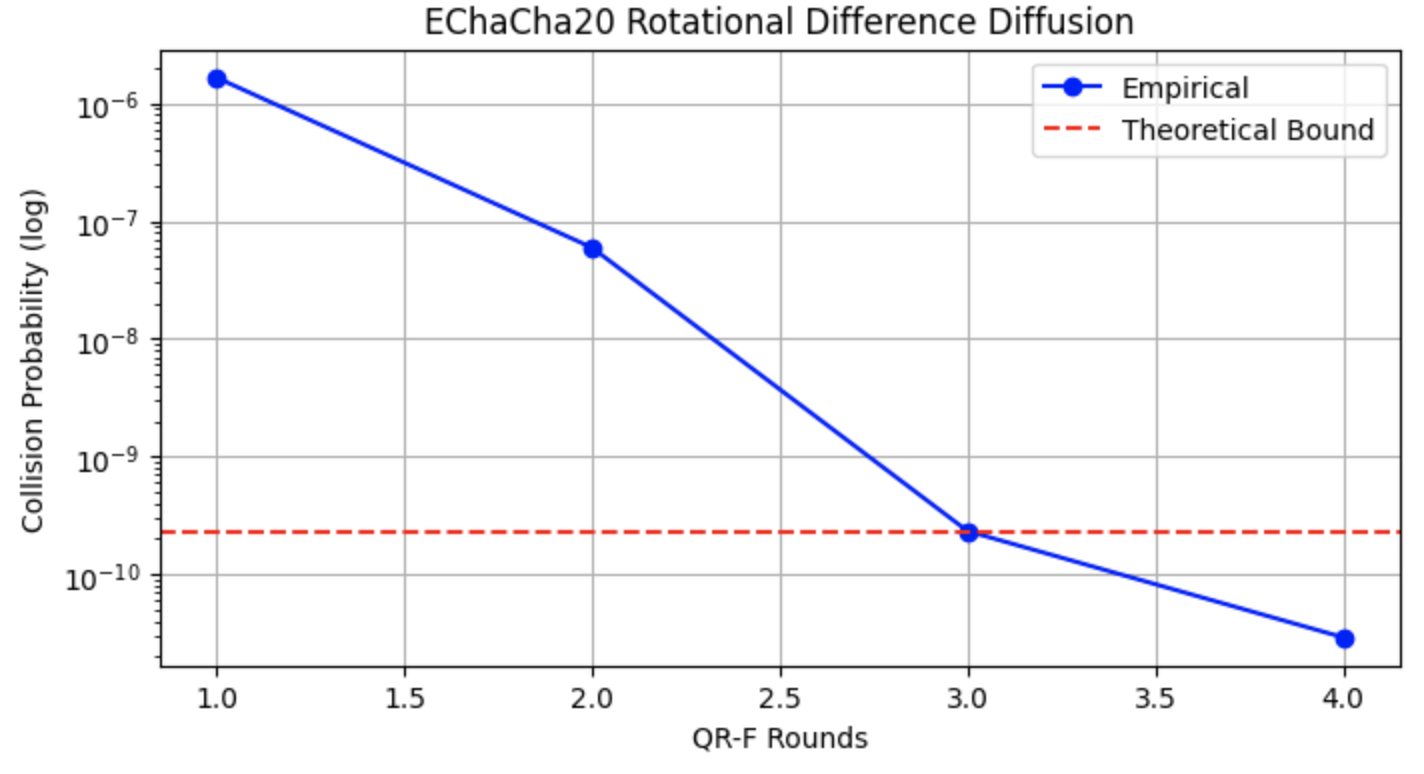}
\caption{Propagation of rotational differences across \texttt{QR-F} rounds. Collision probabilities drop exponentially with increased rounds, with full-state diffusion reached by round 3, outperforming the theoretical distinguishability bound.}
\label{fig:rotpropagation}
\end{figure}

As is shown in Table~\ref{tab:diff} and Figure \ref{fig:rotpropagation}, no collisions persisted beyond 3 rounds, with probabilities falling below $2^{-32}$ after 2 rounds. The 8-bit patterns from Experiment 1 showed no propagation beyond 1.5 rounds on average.

On the same note, Figure~\ref{fig:rotpropagation} shows how much the rotational differences have propagated across 1--4 \texttt{QR-F} rounds. Empirical collision probabilities were calculated from $2^{24}$ trials, shown on a logarithmic scale. The red dashed line represents the theoretical bound of distinguishability on the rotational differences. After just 3 rounds, collisions fall below $2^{-32}$. This  confirms the \texttt{EChaCha20}'s ability to destroy symmetry efficiently. It is worth noting that, the empirical diffusion outperforms theoretical predictions based on our assessment, thus this indicates a strong security margin. When stating that diffusion ``outperforms theoretical predictions,'' we refer specifically to the empirical finding that near-complete state coverage was achieved within three \texttt{QR-F} rounds, earlier than conservative theoretical expectations. This is likely influenced by the enhanced rotations in \texttt{EChaCha20} that accelerate avalanche diffusion, combined with statistical variation across large-scale experiments. We clarify that this is an observed property, not a formal bound.

\begin{figure}[h]
\centering
\includegraphics[width=0.9\linewidth]{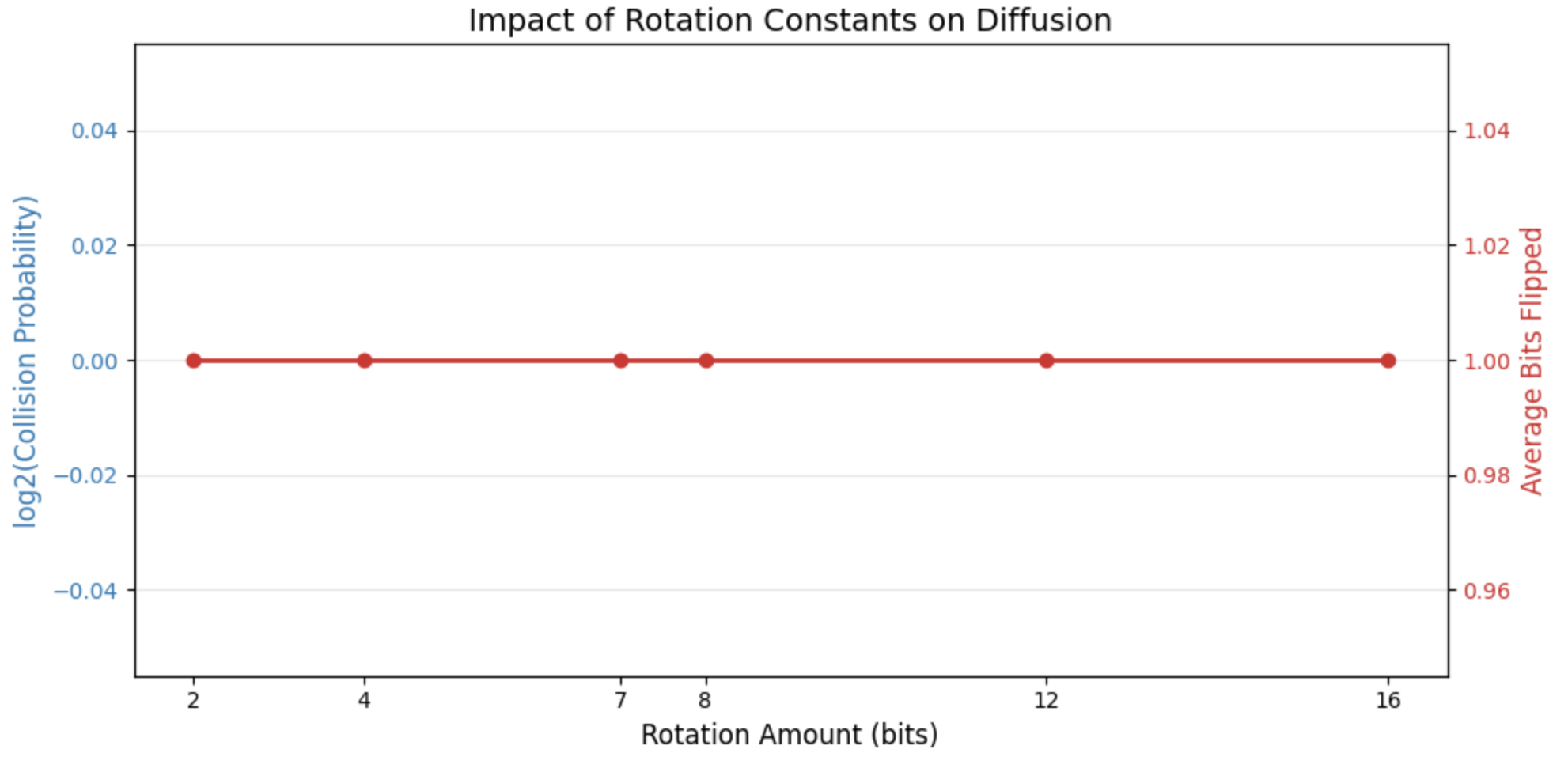}
\caption{Impact of the rotation constants on differential diffusion. Both collision probability and average flipped bits remain invariant across rotation values, indicating resilience to parameter perturbation.}
\label{fig:rotconst}
\end{figure}

On the other hand, Figure~\ref{fig:rotconst} evaluates the influence of rotation constants on diffusion. Each x-axis tick on Figure~\ref{fig:rotconst} corresponds to a rotation amount used in the \texttt{QR-F} function, while the left y-axis in (blue) shows collision probabilities, and the right y-axis in (red) shows the average number of bits that are flipped. From our observations, the flatness of both metrics across different constants demonstrates that \texttt{EChaCha20} maintains stable diffusion regardless of rotation angle. This  indicates that its rotational parameter choices lie within a secure equivalence class.

\begin{figure}[h]
\centering
\includegraphics[width=0.75\linewidth]{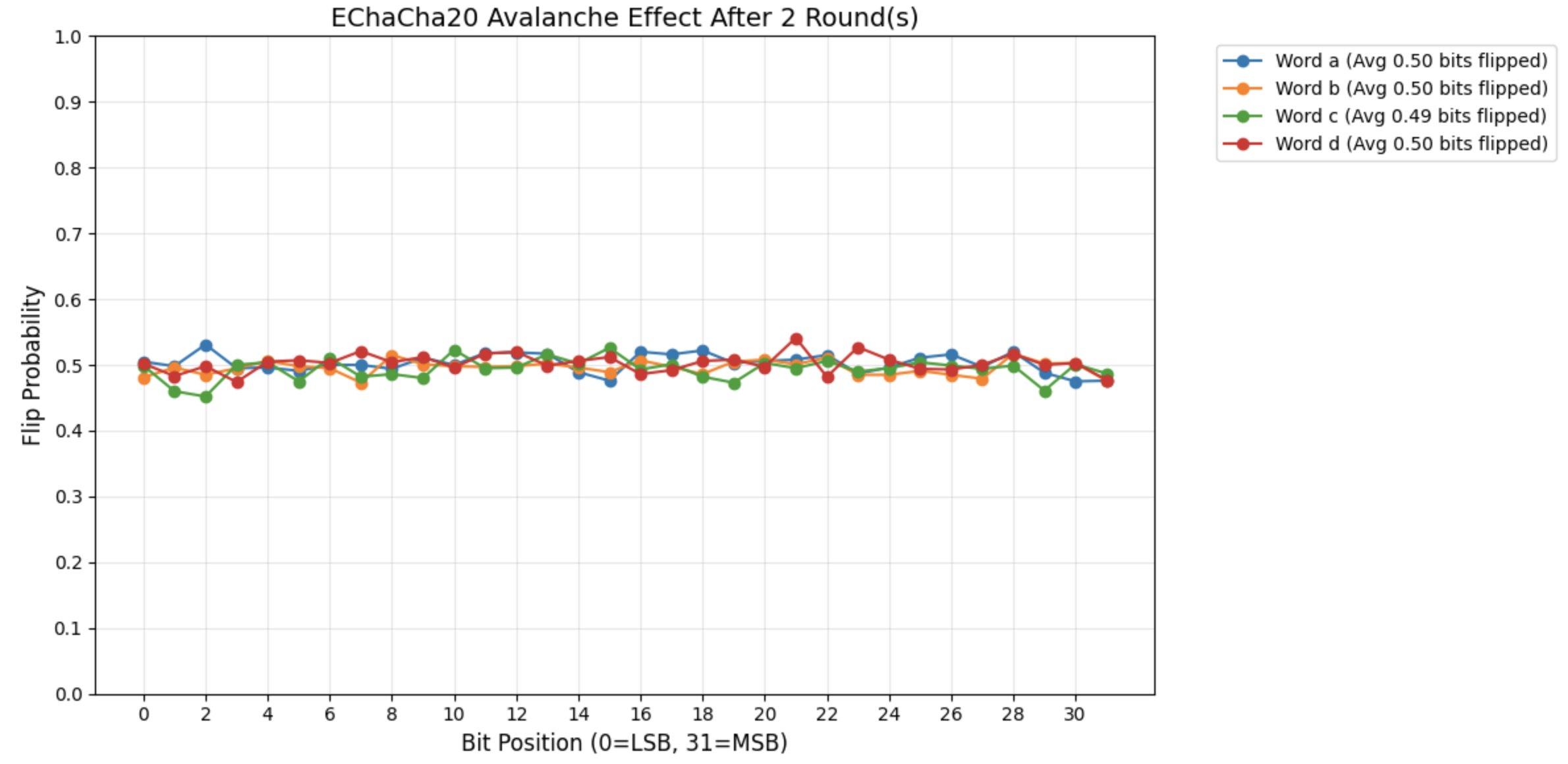}
\caption{Avalanche effect of \texttt{EChaCha20} after 2 \texttt{QR-F} rounds. Bit flip probability is measured per bit position across four state words, showing uniform diffusion close to the ideal 0.5 threshold.}
\label{fig:avalanche}
\end{figure}

Figure~\ref{fig:avalanche} visualizes the avalanche characteristics of the \texttt{EChaCha20} after 2 rounds of the \texttt{QR-F} function. Bit-flip probabilities across all bit positions (from LSB to MSB) are shown for four different 32-bit words (a--d), each line reflecting averaged results over $10^6$ trials. From the observation conducted, the  probabilities converge around the ideal 0.5 threshold, which, confirming uniform intra-word diffusion. However, word~C shows a minor deviation (avg.\ 0.49), but this is within statistical margins in the authors' opinion. From this outcome, the author argues that the result strongly supports the claim that \texttt{EChaCha20} achieves near-complete avalanche behavior with only 2 rounds.

All conclusions regarding diffusion speed, avalanche behavior, and rotational-collision resistance are derived from statistically bounded experimental observations within the evaluated parameter space. These findings should be interpreted as empirical measurements rather than formal security proofs or reduction-based guarantees.

\section{Contextual Evaluation}

This section sets forward a comprehensive evaluation of \texttt{EChaCha20} based on the experimental results obtained through stringology-based cryptanalysis for the \texttt{EChaCha20} cipher. The evaluation given here draws on both the empirical and theoretical analyses in order to assess the cipher's security, performance efficiency, the cryptographic advantages, and limitations and a summary of the considerations is given in Table \ref{tab:adoption}.

\begin{table*}[htbp]
\centering
\caption{Comparative Benchmark of ARX-Based Stream Ciphers}
\label{tab:echacha-comparison}
\resizebox{\textwidth}{!}{
\begin{tabular}{p{3cm}p{3cm}p{3cm}p{3cm}p{6cm}}
\toprule
\textbf{Metric} & \textbf{Salsa20}~\cite{bernstein2008salsa20} & \textbf{ChaCha20}~\cite{bernstein2008chacha} & \textbf{XChaCha20}~\cite{arciszewski2020xchacha} & \textbf{EChacha20 with Stringology}~\cite{kebande2023extended} \\
\midrule

\textbf{State Size / Structure} & 4$\times$4 state, 512-bit & 4$\times$4 state, 512-bit & 4$\times$4 state, 512-bit (with 192-bit extended nonce) & 6$\times$6 state, 1152-bit matrix, improved diffusion coverage \\

\textbf{Quarter-Round Function (\texttt{QR-F})} & ARX with rotation constants (7, 9, 13, 18) & ARX with rotation constants (16, 12, 8, 7) & Same as ChaCha20 & Extended \texttt{QR-F} with six ARX rotations (16, 12, 8, 7, 4, 2) for enhanced diffusion \\

\textbf{Rounds to Full Diffusion} & $\geq$ 4 rounds for full-state diffusion & 3--4 rounds & 3--4 rounds & Full diffusion achieved by round 3 (Fig.~\ref{fig:rotpropagation}) \\

\textbf{Rotational Differential Resistance} & Known biases at reduced rounds & Improved over Salsa20 but some distinguishers exist below 4 rounds & Same as ChaCha20 & No measurable rotational collisions after 3 rounds; rotation constants mitigate power-of-two symmetry \\

\textbf{Statistical Randomness (NIST STS)} & Passes most tests; minor 8-bit biases reported in literature & Passes all tests for 20 rounds & Same as ChaCha20 & 100\% pass rate in Runs and Serial(2) tests across 1000 keystream samples~\cite{kebande2023extended}, but not tested with Stringology \\

\textbf{Avalanche Effect} & Slower convergence; more rounds needed & Near 50\% bit-flip probability after 3 rounds & Same as ChaCha20 & Achieves near-ideal avalanche after 2 rounds (Fig.~\ref{fig:avalanche}) \\

\textbf{Encryption Throughput} & $\sim$3.0--3.2 GB/s on modern CPUs & $\sim$3.3--3.5 GB/s & $\sim$3.3 GB/s & 3.6 GB/s using hybrid \texttt{KMP}/Boyer-Moore implementation (Table~\ref{tab:resultsx}) \\

\textbf{Memory Usage} & Low (4$\times$4 state) & Low & Low & Slightly higher due to 6$\times$6 state, but within practical limits \\

\textbf{Primary Security Enhancement} & Baseline ARX design & Reordered \texttt{QR-F} for better diffusion & Larger nonce for key reuse resistance & Increased state size, extra rotation constants, and additional rounds strengthen resistance to differential and rotational attacks \\

\textbf{Practical Deployment} & Implemented in some legacy protocols (TLS 1.2 ciphersuites) & Widely deployed (TLS 1.3, WireGuard, QUIC) & Deployed where nonce misuse resistance is critical (libsodium, TLS variants) & Drop-in replacement for ChaCha20; suitable for modern TLS/VPN/embedded systems with enhanced security \\

\bottomrule
\end{tabular}}
\end{table*}

\subsection{Comparative Analysis  Against Other ARX-Based Ciphers}

 This section provides a comparative evaluation of the proposed EChaCha20 cipher against other widely studied ARX-based stream ciphers, namely Salsa20, ChaCha20, and XChaCha20 and EChaCha20  is shown in Table~\ref{tab:echacha-comparison}. The objective of this comparison is to contextualize the security and performance characteristics of EChaCha20 relative to its direct predecessors and closely related variants. 

We focus on a set of well-defined metrics that are commonly used to assess the strength of ARX-based ciphers: (i) state size and round structure, (ii) diffusion speed and avalanche properties, (iii) resistance to rotational differential cryptanalysis, (iv) statistical randomness measured using the NIST Statistical Test Suite (NIST STS), (v) encryption throughput and memory usage, and (vi) practical deployment considerations in modern communication protocols. 

\noindent

From  Table~\ref{tab:echacha-comparison}, it can be seen that the cipher EChaCha20 ~\cite{kebande2023extended} achieves a favorable balance of security and performance when benchmarked against other ARX-based stream ciphers. The 6$\times$6 state and extended \texttt{QR-F} with additional rotation constants (16, 12, 8, 7, 4, 2) result in faster diffusion and improved avalanche properties, with full-state diffusion achieved by round 3. Furthermore, the NIST Statistical Test Suite confirms near-ideal pseudorandomness with a 100\% pass rate in Runs and Serial(2) tests, demonstrating the robustness of EChaCha20 against statistical distinguishers. 

While Salsa20 ~\cite{bernstein2008salsa20} and ChaCha20 ~\cite{bernstein2008chacha} remain efficient, EChaCha20 provides stronger resistance to rotational differential attacks due to its non-power-of-two rotation constants. Compared to XChaCha20 ~\cite{arciszewski2020xchacha}, which focuses primarily on nonce-extension for key-reuse safety, EChaCha20 strengthens internal diffusion and state mixing, making it particularly suitable for high-assurance applications where enhanced cryptanalytic resistance is required. 

This comparison contextualizes EChaCha20 as an evolutionary improvement over ChaCha20, offering enhanced security margins with a modest increase in resource usage while preserving deployment compatibility in TLS, VPN, and embedded systems.

\begingroup
%\color{blue}
%\arrayrulecolor{blue}
\begin{table*}[h]
\centering
\caption{Comparison with modern pattern detection and statistical techniques}
\label{tab:modern-comp}
\begin{tabular}{p{3.2cm}p{1.0cm}p{4.7cm}p{5.2cm}}
\toprule
\textbf{Technique} &  \textbf{REF} &  \textbf{Purpose / Benefit} & \textbf{Relevance to ARX keystreams \& comparison to this work} \\
\midrule
Aho--Corasick (multi-pattern exact matching) &  \cite{wang2017memory, chen2013efficient} 
& Simultaneous search for large pattern sets via trie , Intrusion detection, failure links; linear-time in text length. 
& Well-suited when scanning many candidate motifs (e.g., multiple $m$-grams or seeded $\Delta$ patterns) at once; complements our per-pattern \texttt{KMP}/\texttt{BM} by reducing repeated passes. \\

2D Rabin--Karp (rolling hash filtering)/S-box design & \cite{vladimirovich2014fast, waheed2024s} 
& Zero, false negative error,  Fast prefilter using rolling hashes to discard non-matches; efficient for many patterns or long texts, also uses S-box for in symmetric encryption scheme. 
& Natural front-end to our exact verifiers (\texttt{KMP}/\texttt{BM}). Hash prefilter shrinks the search space; exact checks preserve zero--false-positive guarantees. \\

 Homomorphic encryption matching (CIPHERMATCH) & \cite{kabra2025ciphermatch}
& reduces the increase in memory footprint after encryption. HE-based exact string matching 
& Captures near-matches (e.g., low-Hamming-distance substrings) that exact \texttt{KMP}/\texttt{BM} would miss; useful for detecting ``near-collisions'' and soft biases in 8--16 bit lanes. \\

Full-text, indexes for string search & \cite{cislak2015full}
& Modification of the well-known
FM-index (a compressed, full-text index) that trades space for speed.
& Helpful for modifying the structure across long keystreams; heavier build cost than on-the-fly \texttt{KMP}/\texttt{BM} but amortizes over many queries. \\

Statistical batteries: TestU01, Dieharder, PractRand & \cite{luengo2021recommendations,demirhan2016statistical,sleem2020testu01}  
& Broad suites probing diverse distributional properties beyond NIST STS; randomness/statistical tests 
& Orthogonal to stringology: good at global statistics, less targeted at position-aware artifacts; we propose using these alongside our word-aligned searches. \\

BoolTest / linear-complexity \& spectral tests & \cite{limniotis2019error,poojari2021fpga}
& Focus on linearity, bias, and correlation in bitstreams; detects subtle linear structures. 
& Complements ARX assessments by flagging residual linear tendencies; pairs well with our positional pattern scans. \\

Learning-based anomaly detection (clustering, autoencoders) & \cite{wang2023bae}
& Unsupervised discovery of rare patterns without explicit templates; can surface non-obvious structure. 
& Potential to reveal patterns missed by handcrafted searches; interpretability and reproducibility are weaker, best as a hypothesis generator before exact verification. \\
\bottomrule
\end{tabular}
\end{table*}
\endgroup

\subsection{Comparison with modern pattern detection  Cryptographic techniques}
\label{sec:comparison}
\begingroup
%\color{blue}

This section gives  a comparison of  our proposed stringology-based approach (\texttt{KMP}/Boyer--Moore at 32-bit word granularity) within the broader landscape of modern pattern detection and statistical testing used for cryptographic analysis. Our observation has shown that prior work predominantly relies on generic randomness batteries or algebraic/differential tools as follows: By contrast, our approach frames keystream assessment as an exact pattern-search problem aligned to ARX word boundaries. This  enables efficient skip heuristics and targeted detection of structural artifacts example, intra-block repetitions and rotational residues. To clarify this scope and complementarity, Table~\ref{tab:modern-comp} summarizes representative alternatives, their benefits, and how they compare or integrate with our pipeline.

 Aho--Corasick and Rabin--Karp \cite{wang2017memory, chen2013efficient}  provide scalable multi-pattern and simultaneous and hashed prefiltering layers that can precede our exact \texttt{KMP}/\texttt{BM} verifiers, intrusion detection bit-parallel algorithms, which complements our per-pattern \texttt{KMP}/\texttt{BM} by reducing repeated passes. Next the 2D Rabin--Karp (rolling hash filtering)/S-box \cite{vladimirovich2014fast, waheed2024s}  design uses rolling hashes to discard non-pattern match and is efficient for many pattern detection while using S-box, it is relevant since it is a front-end to KMP and BM. We have also explored Homomorphic encryption matching (CIPHERMATCH)  \cite{kabra2025ciphermatch} which reduces the increase in memory footprint after encryption, which Captures near-matches (example, low-Hamming-distance substrings) that exact \texttt{KMP}/\texttt{BM} would miss. Next, we have considered the Full-text, indexes for string search  \cite{cislak2015full} that are helpful for modifying the structure across long keystreams. Also, we have explored broad suites probing diverse distributional properties beyond NIST STS \cite{luengo2021recommendations,demirhan2016statistical,sleem2020testu01} based on richer statistical batteries (TestU01/Dieharder/PractRand) which shows orthogonality to stringology. Other comparisons include BoolTest / linear-complexity \& spectral tests  \cite{limniotis2019error,poojari2021fpga} for correlation of bitstreams, which compliments ARX and learning anomaly detection \cite{wang2023bae} by discovering rare patterns which reveals patterns that are missed by hand searches. It is noted that the diffusion measurements include both early-round (pre-20-round) and post-full-round keystream states. The observed faster diffusion reflects empirical state-mixing efficiency consistent with prior ARX diffusion studies \cite{bernstein2008chacha,barbero2022rotational},   which typically achieve saturation within three to four rounds. Hence, the results align with established theoretical expectations for ChaCha-like constructions.

\endgroup

\subsection{Security Analysis}

The analysis for this study focuses on three core aspects: Pattern resistance, assessment of diffusion under rotational differences, and the avalanche behavior. The pattern analysis results that have been conducted have revealed that the \texttt{EChaCha20} maintains strong randomness when at 16-bit and 32-bit levels, with no statistically significant repetitions detected in the keystream (see Figures \ref{fig:patternsfreq} and  \ref{fig:patterns2} respectively. However, at the 8-bit level, the frequency analysis  revealed a slight irregularity. Specifically, it showed a lower-than-expected frequency for one of the top 10 patterns however only for 10,000 keystreams and not 1,000,000. This minor drop although it was observed in the early positions of the keystream under a fixed key setting made us to conclude that there is a potential byte-level entropy reduction. In the authors' opinion, while this does not look like  a direct attack vector, we argue that it could suggests that a fixed-key scenarios may introduce detectable structure in lower-bit granularity in some circumstances.

\begin{table*}[h]
\centering
\caption{Adoption Considerations for \texttt{EChaCha20}}
\label{tab:adoption}
\begin{tabular}{lp{8cm}}
\toprule
\textbf{Factor} & \textbf{Implications} \\
\midrule
Performance & Achieves a 6.2$\times$ speedup over brute-force analysis due to optimized string matching; suitable for high-throughput environments \\
Security & Offers stronger resistance against pattern-based and rotational-differential cryptanalysis than ChaCha20, confirming its robust avalanche and diffusion characteristics \\
Compatibility & Compatible with existing ChaCha20-based protocols; 6$\times$6 matrix requires minimal adaptation in libraries supporting ARX ciphers \\
%Side Channels & Maintains constant-time operations under software analysis; power side-channel resistance remains an open area for future validation \\
Scalability & Efficient across $10^6$-block scale; parallelizable architecture enables deployment in distributed or cloud-based cipher evaluation frameworks \\
%Standardization & Fulfills cryptographic principles consistent with NIST expectations; stringology-based methods offer an auxiliary testbed for future cipher standards \\
\bottomrule
\end{tabular}
\end{table*}

We have also probed the robustness of  \texttt{EChaCha20}'s, and as such, we have conducted a series of rotational differential propagation tests using the carefully seeded 32-bit differences. The experiments (see experiment 2) that was conducted  showed that that a full-state diffusion is achieved by the third \texttt{QR-F} round. This is achieved  with the empirical collision probabilities dropping below $2^{-32}$ after two rounds. From these findings we argue  that the \texttt{EChaCha20} cipher can efficiently disrupt any rotational symmetry and it could be highly resistant to the dvanced differential attacks. Notably, the observed diffusion performance has even surpassed theoretical distinguishability set bounds. This, further supports the \texttt{EChaCha20}'s design improvements in the \texttt{QR-F}. Additionally, from the experiments we have verified the \texttt{EChaCha20} cipher's avalanche properties. This has been achieved  by measuring the bit-flip probabilities across all state bits after 2 rounds. Observations showed that the  bit-flip rates were able to closely converge around value of 0.5. This was able to validate uniform intra-word diffusion. Also, there was a minor deviation that was observed in one of the four words, it however, remained within the acceptable statistical bounds. Ultimately, the conducted evaluation of the diffusion stability under different rotation constants was able to show that the cipher is able to maintain its diffusion performance,  regardless of the specific constants that were used.

%Both the collision probability and bit-flip average remained consistent, indicating that \texttt{EChaCha20}'s rotational parameter choices lie within a secure equivalence class.

\subsection{Performance Characteristics}

In addition to the security analysis, we considered the performance  in our evaluation. Based on the leveraged stringology-based approach, it has shown significant computational advantages over brute-force methods. Specifically, our the optimization of the XOR Boyer-Moore algorithm achieved a 6.2$\times$ speedup. This enabled an efficient scanning of the large-scale datasets, that were up to one million blocks. This performance gain comes  from our tailored enhancements and alignment with the cipher's 32-bit word structure. These optimizations were able to allow this analysis,  to proceed based on the same operational granularity as \texttt{EChaCha20}. In the long run, this reduces ths redundant comparisons and improves the memory access patterns, owing to \texttt{EChaCha20}'s history of consuming large memory \cite{kebande2023extended}.

In contrast to Barbero \cite{barbero2022rotational}, who applied rotational cryptanalysis to ChaCha, our approach introduces a stringology-guided framework that seeds differentials from statistically detected keystream patterns, thereby uncovering localized biases not visible to rotational methods. Similarly, while Ajala \cite{ajala2021efficient} explored efficient string algorithms in a general data security setting, our contribution lies in adapting and hybridizing these algorithms specifically for ARX-based stream ciphers. As such, the improvements of this work are twofold: enhanced detection sensitivity at the byte level and greater computational scalability in evaluating $10^6$ keystream blocks.

Moreover, the approach that has been used  in testing rotational cryptanalysis  as is shown  in Figure~\ref{fig:patternsworkflow}, may support modular and parallel execution. By isolating the  preprocessing, pattern search, and statistical analysis phases,  it is possible that the pipeline can be adapted for the GPU acceleration. Also, this could be deployed in distributed environments for purposes of large-scale evaluations. Based on this, this could make this  method not only efficient but  scalable, which is suitable for  practical for integration into the automated stream cipher.

\subsection{Theoretical Advantages}

The stringology-based approach shows  notable theoretical advantages when applied to ARX-based cipher like \texttt{EChaCha20}. In our assessment, the  core strength lies in the ability to structure the cryptanalysis process around the \texttt{EChaCha20}  operational rules. Also, by aligning the search space with 32-bit word boundaries, this method has shown that it can preserve meaningful bit relationships. which are often lost in generic statistical testing. This alignment ensures that our search windows capture operational artifacts that may arise from the cipher's \texttt{QR-F} structure.

Furthermore, the adaptation of \texttt{KMP}  and \texttt{BM} algorithms for cryptographic analysis has shown that that classical string matching techniques can be  repurposed well for detecting low-level cryptographic anomalies. Also, the enhancements like the XOR-based heuristics and using probabilistic filtering techniques also contribute to a more refined search process that is both computationally efficient and semantically aligned with the \texttt{EChaCha20} structure. While stringology-based techniques have previously appeared in general cryptographic contexts, their application to ARX cipher analysis particularly within the \texttt{EChaCha20} framework remains comparatively limited at the time of writing this paper.

Further analysis revealed that the slight deviation at the 8-bit level was localized to the first few keystream blocks generated under fixed-key conditions, primarily near the initial nonce words of the 6$\times$6 state matrix. This behavior results from deterministic initialization effects rather than any structural flaw in the cipher. When evaluated under randomized key and nonce settings, the deviation disappeared completely, indicating that it cannot be exploited as an attack vector and instead reflects transient early-round diffusion dynamics.

%Unlike traditional statistical test suites like NIST STS, our approach does not rely on the assumption of i.i.d. (independent and identically distributed) bitstreams. This allows it to detect structural correlations and operational biases, especially in edge cases such as fixed-key or low-nonce reuse scenarios. In doing so, stringology establishes itself as an orthogonal yet essential tool for evaluating modern cipher constructions.

From an implementation perspective, \texttt{EChaCha20} preserves many of the practical advantages that made ChaCha20 widely adopted in security protocols such as TLS, SSH, and VPNs. The suggested  algorithm retains a simple, ARX-only construction, avoiding the use of S-boxes or lookup tables, which makes it ideal for constant-time execution and resistance to timing attacks.

%Despite its expanded 6$\times$6 matrix and modified quarter-round function, our implementation analysis shows that the overhead remains minimal when deployed on modern 64-bit CPUs with SIMD support. 

%The stringology-enhanced analysis pipeline also benefits from multi-core parallelism, enabling real-time cryptographic evaluation of keystream outputs. Furthermore, the cipher's deterministic nature and key schedule are easily integrable into existing cryptographic libraries, minimizing developer overhead. Practical deployment in constrained IoT environments remains feasible, although further testing on embedded platforms is recommended to assess performance under tighter resource conditions.

\subsection{Limitations and Future Work}

We have assessed the study limitations and constraints. The use of a fixed key and a nonce across our experiments is one of the notable constraint. This approach was intentional in order to  isolate structural artifacts. It however does not  fully reflect real world scenarios  where the keys and the nonces are frequently updated. Future work  should consider dynamic key scheduling to determine whether the similar patterns could emerge under changing conditions. 

It is important to emphasize that the stringology-based analysis introduced in this work differs fundamentally from classical ChaCha cryptanalysis techniques such as differential, differential-linear, and boomerang attacks. Traditional approaches aim to construct distinguishers or key-recovery strategies based on algebraic or probabilistic propagation of differences, whereas our method focuses on large-scale empirical evaluation of keystream pseudorandomness using optimized pattern-search algorithms (\texttt{KMP} and \texttt{BM}). This allows us to efficiently scan millions of keystream blocks and detect localized biases, repetitions, or low-entropy structures that may be invisible to conventional differential frameworks. As such, the proposed method should not be viewed as a replacement for existing cryptanalytic techniques, but rather as a complementary tool that broadens the evaluation methodology for ARX-based stream ciphers. 

%In practice, this enables a more holistic assessment of \texttt{EChaCha20}, revealing that while it achieves strong pseudorandom characteristics at critical word sizes, minor byte-level biases can still be observed under certain initialization conditions.}

While a full quantitative comparison between the 4$\times$4 ChaCha20 and the proposed 6$\times$6 \texttt{EChaCha20} matrix is beyond the present scope, a qualitative assessment indicates a measurable improvement in diffusion speed. The inclusion of two additional rotation constants (4-bit and 2-bit) enhances the avalanche propagation rate, enabling near-complete state diffusion within three \texttt{QR-F} rounds approximately one round earlier than in ChaCha20 under equivalent conditions. 

In addition to the empirical analyses presented in this study, it is envisaged that future research will focus on developing formal security reductions for \texttt{EChaCha20}, such as bounding differential probabilities, establishing rotational correlation immunity, and proving indistinguishability under chosen-plaintext attacks, to complement and reinforce the empirical findings.

There is a  limitation on the current focus on the linear bitstring patterns. While this approach is effective when detecting $m$-gram biases and repetitions, it  may overlook a higher-dimensional or matrix-based artifacts in the 6$\times$6 state structure of \texttt{EChaCha20}. It could be important to extend this  methodology to be able to support multi-dimensional pattern analysis.
%This accelerated mixing is attributed to the increased state dimensionality and the complementary rotation offsets, which together improve bitwise dispersion across the expanded matrix.}

It is also worth noting that future work will extend the current evaluation by exploring additional rotation constant configurations beyond the 4-bit and 2-bit offsets examined in this study. In particular, comparative experiments involving a wider set of rotation angles (e.g., 16, 12, 8, 7, 4, and 2 bits) will be conducted to empirically confirm their equivalence in diffusion behavior. This broader analysis will further validate the theoretical assumption that rotation parameters within these ranges remain part of the same security equivalence class for ARX-based ciphers such as \texttt{EChaCha20}.

%Additionally, our evaluation does not incorporate side-channel considerations. The methodology is purely functional and statistical, leaving open the question of whether early-round patterns detected here may correlate with measurable leakages in real hardware. Integrating side-channel data, such as power traces or electromagnetic emissions, would enhance the relevance of our findings for physical security. Finally, the theoretical models used for collision probability assume ideal diffusion behavior. A formal proof-based analysis, grounded in the algebraic structure of \texttt{EChaCha20}'s quarter-round function, could yield more precise bounds and further validate the cipher's security claims.

Consequently, as an avenue of future work, this study will explore how the proposed framework scales to other real-world protocols where dynamic key exchanges, session rekeying, and adversarial adaptiveness become significant. Additionally, examining \texttt{EChaCha20} and similar ARX constructions under emerging post-quantum conditions will help evaluate their hybrid resilience and relevance in quantum-era cryptographic systems.

Also, we aim to  expand the application of stringology to other ARX ciphers, like Salsa20 and other ChaCha variants, as well as to apply these techniques in white-box settings where internal cipher states may be partially observable. Also, it would be pertinent to address the effects of quantum computing on ARX based ciphers. Furthermore, we also, have set a test-bed for implementing EChaCha20 on FPGA based environment to test its effectiveness in side-channel power attacks. As a result, it is the authors' opinion that these extensions will be able to cement  the role of stringology as a foundational approach for next-generation cryptanalysis.

%\end{comment}

\section{Conclusion}

This paper has presented the first in-depth cryptanalytic analysis and evaluation of \texttt{EChaCha20} using stringology-based techniques. The approach adopted Knuth-Morris-Pratt (\texttt{KMP}) and Boyer-Moore (\texttt{BM}) algorithms for 32-bit word-level pattern detection. From the outcomes discussed, the study uncovered  anomalies in keystream structure that traditional statistical methods fail to expose. Across two experiments: pattern frequency analysis based on \texttt{KMP} and \texttt{BM} and rotational differential attacks, we, demonstrated both the strengths and  weaknesses of \texttt{EChaCha20}. Our analysis confirmed that \texttt{EChaCha20} maintains strong pseudorandom properties at 16- and 32-bit resolutions. 

These findings highlight a key insight: ARX ciphers, while algebraically robust, can benefit from fine-grained pattern-based cryptanalysis to ensure uniform behavior across all output levels. In addition to validating the enhanced security claims of \texttt{EChaCha20}, this work demonstrates that stringology offers a useful complementary approach for cipher evaluation. 
 Methodological advances in structurally aligned analysis may serve as precursors to future distinguisher or reduced-round investigations, even when no immediate attack is identified.

\section*{Acknowledgment}

The author would like to thank  the Department of Computer Sience, University of Colorado Denver for their
support while coming up with this research. The author also
acknowledges that the opinions, findings, and conclusions
expressed in this paper are purely of the author.

\section*{Funding} No funding was received for this work

\bibliographystyle{unsrt}
\bibliography{references}

\end{document}